\documentclass[journal]{IEEEtran}

\ifCLASSINFOpdf

\else

\fi

\usepackage{cite}
\usepackage{amssymb,amsfonts}
\usepackage{graphicx}
\usepackage{textcomp}
\usepackage{xcolor}
\usepackage{algorithm}  
\usepackage{algpseudocode}  
\usepackage{subfigure}
\usepackage{pifont}
\usepackage{ulem}
\makeatletter
\let\NAT@parse\undefined
\makeatother
\usepackage{hyperref} 
\hypersetup{hidelinks} 
\usepackage[tbtags]{amsmath}
\allowdisplaybreaks
\renewcommand{\algorithmicrequire}{\textbf{Input:}}  
\def\BibTeX{{\rm B\kern-.05em{\sc i\kern-.025em b}\kern-.08em
    T\kern-.1667em\lower.7ex\hbox{E}\kern-.125emX}}
\usepackage{soul,color,xcolor}      
\definecolor{myColor}{rgb}{0.8039,0,0}   
\makeatletter
\newcommand*{\new}{\@ifnextchar\bgroup{\new@}{\color{myColor}}}
\newcommand*{\new@}[1]{{\textcolor{myColor}{#1}}}
\makeatother
\newtheorem{definition}{\bf{Definition}}
\begin{document}

\title{Adaptive Biased User Scheduling for Heterogeneous Wireless Federate Learning  Network}

\author{Changxiang Wu,~\IEEEmembership{Student Member,~IEEE,}
        Yijing Ren,~\IEEEmembership{Student Member,~IEEE,}
        Daniel~K.~C.~So,~\IEEEmembership{Senior~Member,~IEEE,} and~Jie Tang,~\IEEEmembership{Senior~Member,~IEEE}
\thanks{ C. Wu, Y. Ren and D. K. So are with the Department of Electrical and Electronic Engineering, The University of Manchester, U.K. (e-mail: changxiang.wu,  yijing.ren, d.so@manchester.ac.uk). J. Tang is with the School of Electronic and Information Engineering, South China University of Technology, China (e-mail: eejtang@scut.edu.cn). An earlier progress version of this paper was presented in part at the IEEE International Conference on Communications (IEEE ICC), Rome, Italy, 2023 \cite{wuchangxiang}.}
}

%
%

\markboth{}%
{Shell \MakeLowercase{\textit{et al.}}: Adaptive Biased User Scheduling for Heterogeneous Wireless Federate Learning Network}
%



\maketitle
\begin{abstract}
Federated Learning (FL) has revolutionized collaborative model training in distributed networks, prioritizing data privacy and communication efficiency. This paper investigates efficient deployment of FL in wireless heterogeneous networks, focusing on strategies to accelerate convergence despite stragglers. The primary objective is to minimize long-term convergence wall-clock time through optimized user scheduling and resource allocation. While stragglers  may introduce delays in a single round, their inclusion can expedite subsequent rounds, particularly when they possess critical information.  
Moreover, balancing single-round duration with the number of cumulative rounds, compounded by dynamic training and transmission conditions, necessitates a novel approach beyond conventional optimization solutions. To tackle these challenges, convergence analysis with respect to adaptive and biased scheduling is derived.  Then, by factoring in real-time system and statistical information, including diverse energy constraints and users' energy harvesting capabilities, a deep reinforcement learning approach, empowered by proximal policy optimization, is employed to adaptively select user sets. For the scheduled users, Lagrangian decomposition is applied to optimize local resource utilization, further enhancing system efficiency. Simulation results validate the effectiveness and robustness of the proposed framework for various FL tasks, demonstrating reduced task time compared to existing benchmarks under various settings.
\end{abstract}


\IEEEpeerreviewmaketitle

\section{Introduction}
\IEEEPARstart{I}{n} today's rapidly evolving technological landscape, the proliferation of abundant data has propelled Machine Learning (ML) applications to the forefront, substantially advancing intelligence across diverse domains \cite{jordan2015machine}. While data growth continues its exponential trajectory 
\cite{rydning2018digitization}, ML algorithms thrive on enriching data volume and bridging data silos to boost performance.  However, traditional centralized data processing approaches face considerable challenges in communication efficiency and privacy preservation when aggregating data from large-scale distributed devices \cite{8016573}. In response to these concerns, Federated Learning (FL) has emerged as an innovative distributed ML paradigm, offering a promising solution for secure and efficient data collection and utilization. 
FL, as conceptualized in \cite{mcmahan2017communication}, exploits distributed user data and computational resources to facilitate server-coordinated local model training. Rather than collecting raw data, FL relies on periodic model communication and aggregation to collaboratively train a global model, thereby safeguarding data privacy while also improving communication efficiency.

Despite the demonstrated superiority of FL, contemporary algorithms like Federated Averaging (FedAvg)  \cite{mcmahan2017communication,li2019convergence}, which aggregate local models from randomly selected users, are particularly vulnerable to the detrimental effects of system heterogeneity \cite{9084352}, where stragglers can markedly extend the computational and communicational phases within a global round.
Additionally, statistical heterogeneity \cite{9084352}, characterized by Non-Independent and Identically Distributed (Non-IID) user data, exacerbates this issue. Excluding stragglers in such scenarios is often impractical, as they may possess unique and critical data which is essential for achieving convergence within fewer rounds.
Mitigating the impact of stragglers while leveraging the most informative ones becomes essential in these circumstances. 
An adaptive scheme employing Importance Sampling (IS) \cite{pmlr-v80-katharopoulos18a, 10197174, 9904868, 10443546, 10525198, pmlr-v151-jee-cho22a} has emerged as a promising strategy to address these challenges, as it prioritizes participants based on their critical contributions to the objective, accounting for both system and statistical heterogeneity.
However, existing unbiased adaptive schemes \cite{9904868, 10443546, 10525198} often exhibit slow error convergence rates \cite{pmlr-v151-jee-cho22a}. Furthermore, reliance on resampling scheduling probabilities during aggregation complicates implementation and reduces robustness, as accurate probability estimation becomes a prerequisite \cite{273723}.
To address these limitations, this paper proposes an approach that optimizes biased user scheduling and resource allocation to accelerate convergence in heterogeneous wireless FL systems.

In the considered scenario, the convergence wall-clock time  is determined by the product of single-round duration and the number of cumulative rounds; two inherently conflicting parameters \cite{chenmingzheconvergencetime}. Balancing these factors becomes particularly challenging due to the unpredictable number of required rounds and the dynamic nature of model training and communication during each round. Furthermore, in practical FL scenarios, such as keyboard suggestions \cite{hard2018federated} and smart home Internet of Things (IoT) devices \cite{li2020review}, users’ Energy Harvesting (EH) capabilities introduce further variability into energy constraints.
Moreover, as the rounds progress, the impact of user scheduling on final convergence grows more prominent \cite{xujie}. 
Thus, a forward-looking strategy is critical for arranging users in each round based on their contributions, particularly for those with constrained and dynamic energy resources. Allocating stragglers to rounds with limited contributions risks prematurely depleting their energy, thereby compromising their ability to contribute meaningfully in subsequent critical rounds. Such requirements  render experience-driven Deep Reinforcement Learning (DRL) an attractive candidate. Driven by these challenges,  this paper aims to minimize the long-term convergence in heterogeneous FL networks by leveraging an intelligent policy maker trained using DRL techniques. To achieve this, pioneering works are initially studied.

\subsection{Related Works}
FL, initially introduced in \cite{mcmahan2017communication}, represents a paradigm shift in distributed ML by emphasizing communication efficiency and data privacy. Since its inception, significant research has focused on enhancing FL's convergence properties. Key studies \cite{li2019convergence, wangshiqiangfederated, wanshuo} have analyzed convergence under conditions such as partial user participation and Non-IID data, providing insights into FL optimization.

Addressing system heterogeneity has been an important aspect of wireless FL optimization \cite{wangshiqiangfederated, yangzhaohui, yaojingjing, xujie, chenmingzheconvergencetime, Nishio, shiwenqi, chenmingzhejointlearning, wanshuo}. Notable strategies include optimizing global aggregation frequency \cite{wangshiqiangfederated}, implementing energy-efficient resource allocation \cite{yangzhaohui}, and developing strategies for single-round resource allocation \cite{yaojingjing}. User scheduling also play a crucial role, such as prioritizing users with large gradients \cite{chenmingzheconvergencetime} and ascending scheduling \cite{xujie}. Other strategies include greedy user selection within a time threshold \cite{Nishio}, continuous user selection until the loss function decreases \cite{shiwenqi}, and joint user selection and resource block allocation scheme \cite{chenmingzhejointlearning}. 
A transmission strategy addressing both system and statistical heterogeneity has been proposed in \cite{wanshuo,luobing}. However, these methods are often limited to user selection uniformly at random or Independent and Identically Distributed (IID) scenarios \cite{10443546}.

To tackle statistical heterogeneity, IS \cite{needell2014stochastic}, adapted from Stochastic Gradient Descent (SGD) optimization has been employed to prioritize user selection in FL based on gradient importance \cite{9904868}. Adaptive scheduling has also been investigated in data-level sampling \cite{10525198}. The study in \cite{10443546} combined adaptive user sampling with resource allocation to minimize convergence wall-clock time, but the reliance on unbiased aggregation diminished the impact of high-contribution participants \cite{pmlr-v151-jee-cho22a}. To overcome these limitations, biased scheduling and aggregation methods have been explored \cite{pmlr-v151-jee-cho22a, 273723}. 
Specifically, the biased scheme proposed in \cite{pmlr-v151-jee-cho22a} achieved improved error convergence rates by employing biased sampling and aggregation strategies. A guided user selection scheme for FL training and testing was proposed with respect to the pre-defined utility \cite{273723}. However, these studies do not fully address the challenges posed by energy constraints and system variability and heterogeneity in wireless FL networks.
 
The dynamic nature of wireless channels complicates traditional FL optimization. Advances in ML have led to the adoption of DRL in wireless FL \cite{zhangjie, wanghao, zhanghangjia, zhanyufengl4l, yinbenshun}. 
Specifically, DRL has been utilized for optimizing batch size \cite{zhangjie}, addressing statistical heterogeneity through Deep Q-Networks (DQN) \cite{wanghao}, and managing system heterogeneity \cite{zhanghangjia}. It has also been applied to computional frequency optimization \cite{zhanyufengl4l} and user scheduling through Deep Deterministic Policy Gradient (DDPG) with user staleness guarantee \cite{yinbenshun}. However, a unified framework that concurrently addresses both system and statistical heterogeneity in managing energy constrained users, with the aim of minimizing long-term runtime, remains largely unexplored.
 
 \subsection{Contributions}
 The main contributions of this paper are as follows:
 \begin{itemize}
     \item Unlike prior studies that focuses on isolated aspects, this paper presents a unified framework that concurrently consider statistical heterogeneity (Non-IID data distribution) and system heterogeneity (energy constraints, channel variability, and computational capabilities), aiming to minimize the long-term convergence wall-clock time.
     \item A realistic system model is developed, capturing the dynamic nature of update and upload processes across rounds. Additionally, users are assumed to operate under constrained energy budgets supplemented by EH mechanisms, reflecting real-world conditions.
     \item An adaptive and biased user scheduling scheme is proposed to prioritize participants with high contributions to convergence, with the number of scheduled users dynamically adjusted across rounds based on each round's dynamic contribution to the overall objective.
     \item A Proximal Policy Optimization (PPO)-based DRL methodology is employed to tackle the system's dynamic unpredictability, with a concentrated focus on minimizing the long-term convergence objective. This adaptive policy efficiently responds to evolving network conditions, ensuring near-optimal user selection decisions are made at each global round.
     \item To guarantee convergence and elucidate the impact of user scheduling on FL performance, a theoretical analysis is conducted. Specifically, an explicit term quantifying the acceleration induced by biased user scheduling in Non-IID scenarios is introduced. Moreover, by relaxing the non-convex assumption, the analysis is extended to accommodate neural network-based FL tasks.
     \item To reduce single-round latency, an Alternating Direction Optimization (ADO) method is employed to iteratively optimize communication and computation phases. For communication, a Lagrangian Decomposition-based Resource Allocation (LDRA) scheme optimizes power and subcarrier usage, while a Low Complexity Resource Allocation (LCRA) approach provides closed-form solutions to reduce the algorithm's computational complexity.
     \item Extensive simulations demonstrate that the proposed scheduling policy achieves faster convergence wall-clock time compared to existing benchmarks. 
     The approach exhibits robustness across diverse Non-IID settings, achieving significant efficiency gains through accelerated performance during the mid-training phase.
 \end{itemize}
 
 The remainder of the paper is organized as follows: Section \ref{system model} delineates the wireless FL system model, while Section \ref{convergence anslysis} establishes the convergence bound and problem formulation.  Section \ref{global DRL} and \ref{local Lagrangian} detail the DRL-based user scheduling and local resource allocation methodologies, respectively. Simulation results and their interpretation are presented in Section \ref{simulation results}, followed by concluding remarks in Section \ref{conclusion}.

\section{System Model} \label{system model}

\subsection{Federated Learning Workflow}
Consider a wireless FL system wherein a central server collaborates with a set $\mathcal{N}$ of $N$ heterogeneous users, each possessing a private dataset $\mathcal{D}_n$ ($n=1, 2, \ldots, N$). The server seeks to optimize a global model $\boldsymbol{w}^*$ that minimizes the cumulative loss across all users, formulated as \cite{mcmahan2017communication}:
    \begin{equation}
        \boldsymbol{w}^* \triangleq\arg\min_{\boldsymbol{w}}\left({\mathbb{E}_n\left[ F_n(\boldsymbol{w})\right]}\right)=\arg\min_{\boldsymbol{w}}\left(\sum_{n=1}^Nq_nF_n(\boldsymbol{w})\right), \label{flobjective}
    \end{equation}
where $q_n \ge 0$, and $\sum_{n=1}^N q_n = 1$ represents the weight of user $n$, and $F_n(\boldsymbol{w}) = D_n^{-1} \sum_{j \in \mathcal{D}_n} f_j(\boldsymbol{w}; \boldsymbol{x}_j, y_j)$ represents the local loss function, with $f_j(\boldsymbol{w}; \boldsymbol{x}_j, y_j)$ evaluating model performance on sample $(\boldsymbol{x}_j, y_j)$. 
The FL workflow proceeds iteratively in global rounds initiated by the central server, which broadcasts the scheduling policy along with the latest global model $\boldsymbol{w}_k$. Then, the selected users download the global model and conduct localized training on their datasets via mini-batch SGD for $E$ epochs, updating the model as follows:
    \begin{equation}    \boldsymbol{w}_{k,n}^{\tau+1}=\boldsymbol{w}_{k,n}^{\tau} - \eta_k\nabla{l_{\mathcal{B}_{k,n}}(\boldsymbol{w}_{k,n}^{\tau})},\quad \tau=0,1 \cdots E-1, \label{localupdate}
    \end{equation}
where $k$ refers to the current global round index; $\tau$ is the local epoch index; $w_{k,n}^{\tau+1}$ depicts the updated model in the $(\tau+1)$-th epoch, with $\boldsymbol{w}_{k,n}^0=\boldsymbol{w}_k$  at the start of each global round;  the learning rate is denoted as $\eta_k$; $\nabla{l_{\mathcal{B}_{k,n}}(\boldsymbol{w}_{k,n}^\tau}) = \frac{1}{|\mathcal{B}_{k,n}|}\sum_{j\in\mathcal{B}_{k,n}}\nabla f(\boldsymbol{w}_{k,n}^{\tau};\boldsymbol{x}_{j},y_{j})$ stands for the mini-batch stochastic gradient. 
The local updating process incurs noise due to the randomly selected batch $\mathcal{B}_{k,n}$ in every update, with unbiased expectation $\mathbb{E}_{\mathcal{B}_{k,n}}({\nabla l_{\mathcal{B}_{k,n}}(\boldsymbol{w}}))=\nabla F_n(\boldsymbol{w})$, introducing dynamic stochasticity in model updates during each global round. Upon completing $E$ local epochs, users upload their updated models to the server, which facilitates aggregation through the following process:
    \begin{equation}
        \boldsymbol{w}_{k+1} =\frac{1}{V_k}{\sum_{n\in\mathcal{V}_k} \boldsymbol{w}_{k,n}^E}, \label{aggregation}
    \end{equation}
where $\mathcal{V}_k$ represents the set of participating users in round $k$, and $V_k$ is the cardinality of $\mathcal{V}_k$. Unlike other adaptive unbiased methods that incorporate normalization, the aggregation here is a straightforward linear sum, as the scheduling probabilities are integrated during the user selection process \cite{pmlr-v151-jee-cho22a}. This process continues until the predefined target is achieved.
\subsection{Enabling Federated Learning over Wireless Network}
In FL networks, selected users perform training tasks utilizing their computation and communication modules, which incurs energy and time costs. These tasks are handled by computation modules, typically Central Processing Units (CPUs), which perform local model training. The computation time for user $n$ during one global round is defined as: 
    \begin{equation} \label{computation time}
        t_{k,n}^\text{(cp)} =  {E c_n d_n}/{f_{k,n}}, \quad \forall n\in{\mathcal{N}},
    \end{equation}
where $c_n$ indicates the number of CPUs cycles needed to process one bit of data \cite{yangzhaohui}, $d_n$ refers to the total bits per mini-batch, and $f_{k,n}$ symbolizes the adjustable CPUs frequency to meet task requirements \cite{zhanyufengl4l}. The energy consumed per CPUs cycle follows $\kappa f_{k,n}^2$, where $\kappa$ is a parameter related to chip architecture, indicating the switched capacitance during computation \cite{yangzhaohui, yaojingjing}. Consequently, the computation energy for one global round at user $n$ is:
    \begin{equation} \label{computation enerfy}
        e_{k,n}^\text{(cp)}=\kappa E f_{k,n}^2 c_n d_n, \quad \forall n\in{\mathcal{N}}.
    \end{equation}

Once the computation is completed, selected users transmit their models to the Base Station (BS), which hosts the FL central server, via uplink Orthogonal Frequency-Division Multiple Access (OFDMA) channels with fading. Accurate Channel State Information (CSI) is assumed to be available at the BS in each round. The BS can employ up to $M$ subcarriers for all $N$ users. The transmission rate for the link between user $n$ and the BS is given by:
    \begin{equation} 
        R_{k,n} = \sum_{m=1}^MBc_{k,n}^m\log_2(1+P_{k,n}^m \varphi_{k,n}^m), \quad \forall n\in{\mathcal{N}},
    \end{equation}
where  $B$ denotes the bandwidth per subcarrier; $P_{k,n}^m$ refers to the transmission power on subcarrier $m$; $\varphi_{k,n}^m = \frac{h_{k,n}^m}{N_0B}$ is the Channel to Noise Ratio (CNR) for user $n$ on subcarrier $m$ in the $k$-th round and $N_0$ the noise power spectral density. The subcarrier assignment indicator $c_{k,n}^m = 1$ denotes that subcarrier $m$ is allocated to user $n$, and $c_{k,n}^m = 0$ otherwise. Given that local models share a uniform structure, their corresponding package size $\Pi$ is assumed to be the same.  Thus the communication time of user $n$  can be expressed as:
    \begin{equation} \label{communication time}
            t_{k,n}^\text{(cm)} = {\Pi}/{R_{k,n}}, \quad \forall n\in{\mathcal{N}}.
    \end{equation} 
Accordingly, the required energy of user $n$ for uploading  is:
    \begin{equation} \label{communication energy}
        e_{k,n}^\text{(cm)} = \sum_{m=1}^M P_{k,n}^m c_{k,n}^m t_{k,n}^\text{(cm)}, \quad \forall n\in{\mathcal{N}}.
    \end{equation}
In this work, it is assumed that the participants are equipped with EH facilities and can harvest energy from outside sources like radio frequency EH \cite{6951347}. Therefore, the energy budget of user $n$ at the beginning of round $k$ can be calculated as:
\begin{equation}
    E_{k,n} = \min\big\{ E_{k-1,n} - e_{k-1,n}^\text{(cp)} - e_{k-1,n}^\text{(cm)} + E_{k-1}^\text{HE}, E_{\text{max}}\big\},
\end{equation}
where $E_{k-1}^\text{HE}$ denotes the harvested energy from the outside source in the previous round, which is assumed to follow a Poisson process with an expectation value $\varkappa$, and $E_{\text{max}}$ refers to the maximum battery capacity of each user.
In a synchronous FL framework, the server must await the slowest (straggler) user's model update before aggregation can proceed, defining the completion time for a round as:
    \begin{equation}
        t_k = \max_{n\in\mathcal{V}_k} \left\{t_{k,n} ^\text{(cp)} +  t_{k,n} ^\text{(cm)} \right\}. 
    \label{time}
    \end{equation}
The analysis excludes downlink transmission delays, considering that the BS has exclusive access to abundant communication resources \cite{chenmingzhejointlearning, wanshuo, wangshiqiangfederated, yangzhaohui, luobing, yaojingjing,Nishio,shiwenqi}. Given the aforementioned resource model, an integrated resource allocation strategy that optimizes both computational and communication capabilities is crucial for minimizing latency per round for selected users. Furthermore, the quality and quantity of selected users influence both the latency of specific rounds and the total number of rounds required for convergence. Therefore, the impact of user scheduling on convergence will be explored through theoretical analysis in the subsequent sections.

\section{Convergence Analysis and Problem Formulation} \label{convergence anslysis}
\subsection{Convergence Analysis}
This subsection conducts a convergence analysis under the Non-IID data setting to theoretically support the proposed adaptive biased user scheduling strategy and evaluate its influence on the convergence process. To facilitate a comprehensive understanding, following assumptions are introduced.  Notably, the convexity assumption commonly adopted in existing studies is relaxed, enabling the analysis to accommodate typical FL scenarios involving non-convex  neural networks.
\newtheorem{assumption}{\bf{Assumption}}
\begin{assumption} \label{assumption1}
    The loss function $F$ is continuously differentiable and $\beta_k$-smooth, meaning that for any $\boldsymbol{w}$ and $\boldsymbol{v}$:
    \begin{equation}
        \left\| \nabla F(\boldsymbol{w}) - \nabla F(\boldsymbol{v}) \right\| \leq \beta_k \| \boldsymbol{w} - \boldsymbol{v} \|.       
    \end{equation}
    It should be noted that the gradient Lipschitz smoothness parameter $\beta_k$ is allowed to vary as FL training proceeds, since an invariant gradient Lipschitz value does not exist.
\end{assumption}
\begin{assumption} \label{assumption2}
The variance of the SGD gradients on each user is bounded by:
    \begin{equation}
        \mathbb{E}\left\|\nabla F_n(\boldsymbol{w}_{k,n}^{\tau})-\nabla l_{\mathcal{B}_{k,n}}(\boldsymbol{w}_{k,n}^\tau) \right\|^2 \leq \sigma.
    \end{equation}
\end{assumption}
\begin{assumption}\label{assumption3}
    The divergence between local and global gradient norms is mathematically bounded as follows \cite{9556559,9759241}:
    \begin{equation}
        \frac{1}{N}\sum\nolimits_{n=1}^N\left\| \nabla F_n(\boldsymbol{w}) - \nabla F(\boldsymbol{w})\right\|^2 \leq \alpha.
    \end{equation}
\end{assumption}
 Similar to \cite{9556559,9759241}, the gradient norm divergence is adopted to model the Non-IID nature of the data distribution, where larger value of $\alpha$ indicates higher level of data heterogeneity for user $n$. To further assist the convergence analysis, the selection bias is introduced.
\begin{definition} \label{definition}
    Motivated by the definition of selection skew in \cite{pmlr-v151-jee-cho22a}, the user selection bias for any given user scheduling scheme $\pi_\mathcal{S}$ is defined as:
    \begin{equation}
       \rho =  \frac{\mathbb{E}_{\mathcal{S}_k} \left[\sum_{n\in\mathcal{V}_k}\mathcal{G}_n  \left\| \nabla F_n(\boldsymbol{w}_k^n) - \nabla F(\Bar{\boldsymbol{w}}_k) \right\|^2\right]}
       {\sum_{n=1}^{N}q_n  \left\| \nabla F_n(\boldsymbol{w}_k^n) - \nabla F(\Bar{\boldsymbol{w}}_k) \right\|^2 },
    \end{equation}
    where $\mathcal{G}_n$ is the aggregation weight, which varies depending on the scheduling policy. In this paper, $\mathcal{G}_n\! =\! \frac{1}{V_k}$, and for random or adaptive scheduling with unbiased aggregation, $\mathcal{G}_n\! = \!\frac{q_n}{V_k p_{k}^n}$ \cite{10443546}, with $p_k^n$ denotes the scheduling probability of user $n$ in round $k$. For random or unbiased aggregation schemes, it can be observed that \(\rho = 1\). In contrast, for biased schemes employing IS, \(\rho \geq 1\), as the aggregation does not include probability normalization, allowing the weighted sum in the numerator to increase adaptively \cite{pmlr-v151-jee-cho22a}.
    Based on these assumptions and definitions, the following theorem is derived.
\end{definition}
\newtheorem{theorem}{\bf{Theorem}}
\begin{theorem} \label{theorem1}Assuming a decaying learning rate $\eta_k \leq \frac{1}{\beta_k}$, and $G_k\,=\, \max_{n\in\mathcal{V}_k}\left\|  \nabla F_n({\boldsymbol{w}_k^n}) \right\|^2$, the expected gradient norm of the adaptive biased scheduling is bounded by:
   \begin{align} \label{convergence bound}
         &\mathbb{E}\left[\frac{1}{\Omega_\epsilon}\sum_{k=0}^{\Omega_\epsilon-1} \left\|\nabla F(\Bar{\boldsymbol{w}}_{k}) \right\|^2\right] \nonumber\\
         \leq&   \frac{2}{\eta_\text{min} \Omega_\epsilon}\left[ F(\Bar{\boldsymbol{w}}_{0}) - F^*\right] + \frac{2}{\eta_\text{min}\Omega_\epsilon}\sum_{k=0}^{\Omega_\epsilon-1} (\vartheta_k + \varepsilon_k + \psi_k),\qquad 
         \nonumber\\
        & \text{with:} \nonumber\\
        & \psi_k = \frac{\eta_k \mathcal{E}}{\rho}\alpha, \nonumber\\
        &\vartheta_k = \left( \frac{\eta_k}{2 V_k} + \frac{4 \eta_k E \mathcal{E}}{\rho} \right) \sigma,\nonumber\\
        &\varepsilon_k = \frac{2\eta_k}{V_k} G_k
        \!+\!\frac{\eta_k \mathcal{E} N}{2} {G_k} \!+\! \frac{4 \eta_k \mathcal{E}}{\rho}\!  \sum_{\tau=0}^{E-1}G_{k,\tau}, 
    \end{align}
\end{theorem}    
where $\Omega_\epsilon$ denotes the number of required global rounds to achieve the predefined target accuracy $\epsilon$; $\mathcal{E}=1+\sum_{n=1}^N (p_k^n-q_n)^2$; the first term reflects the initial optimization gap after $\Omega_\epsilon$ rounds; $\psi_k$  quantifies the error introduced by the Non-IID data; $\vartheta_k$ relates to the SGD error caused by subsequent training; $\varepsilon_k$ captures the dynamic gradient deviation arising from heterogeneous data distribution \cite{10525198}.
\begin{IEEEproof}
    See Appendix \ref{appendixA}.
\end{IEEEproof}

Notably, the FL algorithm may converge to a local minimum due to the non-convex nature of the objective function \( F(\boldsymbol{w}) \). Therefore, similar to \cite{10525198,9759241}, the expected gradient norm is employed as a convergence indicator. Specifically, the FL task is considered to achieve a \(\Delta\)-suboptimal solution if the expected gradient norm is less than \(\Delta\). Therefore, the convergence bound ensures the convergence of the FL algorithm to a stationary point. Key insights from the convergence results are as follows: As training progresses, the gradient norm gap diminishes, and the algorithm eventually converges when $\Omega_\epsilon$ becomes sufficiently large. 
The non-linear relationship expressed by the function $\frac{1}{\Omega_\epsilon}$, compounded by the dynamic gradient term $\varepsilon_k$ cause the contribution of each round to vary dynamically, requiring adaptive adjustments in the number of scheduled users based on their relative impact. 
As reflected in $\varepsilon_k$ and $\psi_k$, the scheduling should jointly consider the dynamic gradient Lipschitz and the Non-IID distribution, emphasizing the importance of an adaptive scheduling scheme.
Moreover, in the presence of Non-IID data distributions, additional errors reflected in $\psi_k$ arise, yet these can be mitigated through biased user scheduling, which improves the convergence rate to $\mathcal{O}(\frac{1}{\rho \Omega_\epsilon})$. The above analysis and (\ref{time}) also underscore the intricate interdependence between the user selection pattern, ${p}_k^n$, and the required training rounds, $\Omega_\epsilon$. While increasing the number of scheduled users can reduce $\Omega_\epsilon$, it may prolong single-round durations due to stragglers. Moreover, modifications to $\Omega_\epsilon$ influence the optimal scheduling pattern at each step. These insights highlight the critical need for a holistic, forward-looking approach that jointly optimizes adaptive user scheduling and dynamic resource allocation to minimize long-term convergence time in heterogeneous FL networks.
\subsection{Problem Formulation}
This paper aims to accelerate the FL convergence for heterogeneous networks, in which the user scheduling, subcarrier assignment, CPUs frequency and power allocation are jointly optimized. Formally, the optimization problem is defined as:
   \begin{align} \label{initial problem}
        \textbf{P1}: &\min_{\mathcal{V}_k, c_{k,n}^m, P_{k,n}^m, f_{k,n} } \hfill\ \ \sum\nolimits_{k=0}^{\Omega_\epsilon-1}t_k \hfill&\\
         &\qquad \ \   \text{s.t.} \hfill \ \ \nonumber  c_{k,n}^m \in \left\{ 0,1 \right\},\hfill\ \forall m,\  \forall n\in\mathcal{V}_k\hfill &\tag{16a}\\
           &\qquad \qquad \ \  e_{k,n} ^\text{(cp)} + e_{k,n} ^\text{(cm)} \leq E_{k,n},\hfill\ \forall k,\  \forall n \in\mathcal{V}_k, \hfill \tag{16b}\\
         &\qquad \qquad\ \  \sum\nolimits_{m=1}^Mc_{k,n}^m\!  P_{k,n}^m \leq P_n^\text{max},\forall n\in\mathcal{V}_k, &\tag{16c}\\
         &\qquad \qquad \ \ P_{k,n}^m \geq 0, \hfill\ \forall m,\  \forall n\in\mathcal{V}_k,\ \hfill&\tag{16d}\\
         &\qquad \qquad \ \ \sum\nolimits_{n\in\mathcal{V}_k} c_{k,n}^m = 1, \hfill\  \forall m, \hfill &\tag{16e}\\
         & \qquad \qquad \ \ f_{\text{min}}\leq f_{k,n} \leq f_{\text{max}}, \hfill\  \ \forall n\in\mathcal{V}_k,\hfill &\tag{16f}\\
        & \qquad \qquad \ \ F(\boldsymbol{w}_{\Omega_\epsilon}) - F(\boldsymbol{w}^*) \leq l_\epsilon. \hfill&\tag{16g} 
   \end{align}
 Constraint (\ref{initial problem}b) limits the total energy consumption of each user remains within their dynamic energy budget; constraints (\ref{initial problem}c) and (\ref{initial problem}d) restrict the total allocated power of one user to $P_n^{\text{max}}$ and the allocated power for each subcarrier non-negative, respectively; constraint (\ref{initial problem}e) guarantees each subcarrier is exclusively assigned to one user; constraint (\ref{initial problem}g) ensures the final model loss satisfies the target loss requirement. In this paper, we simulate the FL-aided image recognition task; therefore the loss can be represented by the recognition accuracy $\epsilon$ for convenient. To derive the optimal solution of \textbf{P1}, the following challenges need to be considered:
\begin{itemize}
    \item The problem involves nonlinear computations in the objective function and constraint (16b), coupled with the binary nature of user scheduling and subcarrier assignment. These make \textbf{P1} a non-convex Mixed Integer Non-Linear Programming (MINLP) problem, presenting significant challenges for resolution using traditional methods.
    \item The running mechanism of FL limits the server’s access to raw data, allowing only partial participation and model weight updates in each round. This restricts the ability to derive explicit mathematical formulations for data distributions and gradient calculations, highlighting the necessity for scheduling strategies that leverage partial information and experience-driven optimization.
    \item The dynamism of models, channel states and energy budget in each round, due to the stochastic sampling of SGD, channel fading and EH, necessitates adaptable transmission and training schemes to accommodate these inherent dynamic fluctuations.
    Furthermore, it can be seen from (\ref{computation time})\,-\,(\ref{communication energy}) that expediting computation and communication in one round can be achieved by excluding stragglers. However, the convergence bound (\ref{convergence bound}) highlights that the absence of informative users might potentially extend total rounds required for the whole task completion. 
    Therefore, scheduling strategies must jointly weighs long-term benefits as well as the instantaneous reward to strike a balance to minimize the overall time.
\end{itemize}
Given the above challenges, the experience-driven DRL can be a potential scheme to obtain a near-optimal scheduling solution. Following the DRL-selected user-set, local resource allocation can be designed by traditional optimization tools to further reduce the single-round computation and communication time.

\section{Deep Reinforcement Learning based User Scheduling} \label{global DRL}
This section details the integration of DRL into the wireless FL to adaptively determine the selection of users in each global round. The proposed framework leverages a Markov Decision Process (MDP) to model decision-making process, with the DRL agent trained via the PPO algorithm \cite{PPO}.
\subsection{Markov Decision Process Design}
Based on discussions in the above sections, the state space needs to capture  essential factors influencing the agent decision,  including instantaneous link qualities, model and data distribution states, CPUs frequency configurations, and users' energy budgets, thereby accommodating the system and statistical heterogeneity. Inspired by findings in \cite{zhao2018federated}, which show that model weight divergence effectively captures data diversity and gradients among users, a normalized divergence vector ${\mathcal{W}}_k$ is introduced to represent system heterogeneity. Its update process can be expressed as:
\begin{equation}
    {\mathcal{W}}_k = \left\{ {\|\boldsymbol{w}_{k,n}^E-\boldsymbol{w}_k\|} / {\| \boldsymbol{w}_k \|}\right\}, \hfill \ \forall n\in\mathcal{V}_k,\hfill
\end{equation}
where $\boldsymbol{w}_{k,n}^E$ represents the collected model weights of user $n$, and $\boldsymbol{w}_k$ corresponds to the aggregated model of all scheduled users at the start of $k$-th round. Initially, $\mathcal{W}_0$ is set to zero as the starting model for all users is identical. 
As FL progresses, the divergence vector is updated only for the selected users in each round. To inform the DRL agent about the current FL progress, the gap between the current model accuracy ${\boldsymbol{\epsilon}}_k$ and the target accuracy $\epsilon$, i.e., $\epsilon - \boldsymbol{{\epsilon}}_k$, is included in the state space. Thus, the comprehensive state space is:
\begin{equation} \label{state}
        {\mathcal{S}}_k = \{  \boldsymbol{h}_k, \boldsymbol{f}_{\text{min}}, \boldsymbol{f}_{\text{max}}, \mathcal{W}_k, \epsilon - \boldsymbol{ {\epsilon}}_k, \boldsymbol{e}_k\},
\end{equation}
where $ \boldsymbol{h}_k =\{ h_{k,1}, \cdots, h_{k,N} \}$ refers to the channel quality set acquired by the BS at the start of the $k$-th round; $h_{k,n}={\sum_{m=1}^M h_{k,n}^m}/{M}$, and variables $\boldsymbol{f}_{\text{min}}$ and $\boldsymbol{f}_{\text{max}}$ indicate feasible CPUs frequency ranges. Based on the current state, the DRL agent determines the corresponding action, which includes the number of users to be selected and the user scheduling probabilities. To maintain scalability in large-scale systems, actions are modeled as continuous distributions, thus preventing exponential growth in the action space, which can occur with discrete algorithms. The defined action space is:
\begin{equation} \label{action}
     {\mathcal{A}}_k =\left\{m_k, \boldsymbol {p}_k\right\},
\end{equation}
where $m_k$ represents the fraction of activated users, and $\boldsymbol {p}_k$ encapsulates the user scheduling probabilities. The scheduler then performs the selection by setting the queue length as $\lfloor m_k N \rfloor$, and then admitting users based on the descending order of $\boldsymbol {p}_k$ until the queue reaches its limit. 

Given the overarching objective in \textbf{P1} to minimize the convergence wall-clock time, the reward function in the $k$-th round is designed to be inversely proportional to the single-round completion time, encompassing both training and uploading durations. This alignment ensures that the DRL agent's goal coincides with the primary objective of the paper, and the reward is expressed as follows:
\begin{equation} 
\mathcal{R}_k = -t_k.\label{reward} \end{equation} 
The negative sign motivates the agent to minimize the total convergence time, thereby aiming to reach the target accuracy in the minimum possible duration.

\subsection{Training DRL Agent over Wireless FL Network}
Considering the features of the given problem and MDP, where dynamic and training sample-constrained environments are prevalent, PPO \cite{PPO} is chosen for both pre-offline DRL agent training and subsequent online policy updates for the following reasons: 
\begin{itemize}
    \item PPO is well-suited for handling continuous action spaces, making it ideal for the considered user scheduling policy, which involves determining continuous values for the proportion of selected users and their scheduling probabilities. Moreover, the exploit-exploration mechanism \cite{PPO} guarantees fairness in user scheduling and robustness of the algorithm \cite{273723}.
    \item Considering the highly dynamic characteristics of wireless FL networks, PPO's ability as an online learning algorithm enables adaptive updates to the DRL agent using newly collected state-action trajectories \cite{8303773}. Notably, to prevent delays in the FL process, policy updates are restricted to occur only once the buffer is full and a FL task has been completed.  
    \item PPO operates in an on-policy learning framework, which means it learns from the data it generates during training. Therefore,  PPO becomes a suitable approach in wireless FL, where obtaining off-policy MDP data (data for training the DRL agent is generated by a different policy) in advance can be challenging and unrealistic.
\end{itemize}

 To train the actor network by PPO, in the $k$-th global round, the agent asks for the state $\mathcal{S}_k$. Subsequently, these states are fed into the old actor network $\pi_{\boldsymbol{\theta}_A'}$ to obtain the corresponding transmission decision ${\mathcal{A}}_k$, which is then broadcast by the BS alongside the latest global model $\boldsymbol{w}_k$. Upon receiving the control signal, scheduled participants engage in $E$ local training epochs and subsequently transmit their local models for aggregation. 
 The collected experience tuples $(\mathcal{S}_k, \mathcal{A}_k, \mathcal{R}_k, \mathcal{S}_{k+1})$ are stored in a replay buffer. Once the buffer is full, the PPO optimizer trains the agent by performing mini-batch SGD (line 13-17 of \textbf{Algorithm~\ref{algorithmcompleteTraining}}). The complete algorithm is summarized in \textbf{Algorithm~\ref{algorithmcompleteTraining}}. Notably, since transitions can be stored, the DRL agent can be trained offline without adding delays to FL training \cite{zhanyufengl4l}. After training, the actor network $\pi_{\boldsymbol{\theta}_A}$ can be deployed in a real-world FL scenario.

 In  this work, Deep Neural Networks (DNNs) with two hidden layers, each containing $Z$ neurons, are used for modeling both the actor and critic networks. Each state type is first processed through an intermediate layer before aggregating into a unified hidden layer. The number of users scheduled and scheduling probability are independently produced through separate output layers. Therefore, the training time complexity is $\mathcal{O}(\Psi  Z^2)$, where $\Psi$ represents the required rounds for agent training, and $Z$ is the number of nodes for each hidden layer. Meanwhile, the time complexity for action decision-making per round, when deploying the trained actor network in a real FL scenario, is $\mathcal{O}( Z^2)$.
    \begin{algorithm}[t]  
        \caption{Training DRL Agent with PPO}
        \begin{algorithmic}[1]  
        \Ensure randomly initialize the parameters of the actor network ($\pi_{\boldsymbol{\theta}_A}$) and critic network ($V_{\boldsymbol{\theta}_C}$); initialize an empty replay buffer; set
        $\pi_{\boldsymbol{\theta}_A'} \leftarrow \pi_{\boldsymbol{\theta}_A}$
        \For{each training episode}
            \State Reset the wireless FL environment to its initial  $\mathcal{S}_0$
            \While{one FL task is not done}
                \State Derive the action ${\mathcal{A}}_k$ by feeding the current state \Statex \qquad \quad $\mathcal{S}_k$ into the actor network ($\pi_{\boldsymbol{\theta}_A'}$)
                \State Select user sets $\mathcal{V}_k$ based on derived action ${\mathcal{A}}_k$
                \State The selected users perform local model updates \Statex \qquad \quad according to (\ref{localupdate}) and send back the latest model
                \State Update the global model based on (\ref{aggregation}) and calculate\Statex \qquad \quad the reward $\mathcal{R}_k$ following (\ref{time}) and (\ref{reward})
                \State The DRL agent observes the next state $\mathcal{S}_{k+1}$
                \State Store the transition $(\mathcal{S}_k, \mathcal{A}_k, \mathcal{R}_k, \mathcal{S}_{k+1})$ into buffer
                \State Calculate advantage function for PPO \cite{PPO}
                \State $k = k+1$
            \EndWhile
            \If {buffer is full \textbf{and} during FL idle time}
                \State Calculate the actor loss and the critic loss
                \State Update the actor and critic networks by SGD      
                \State $ \pi_{\boldsymbol{\theta'_A}} \gets\pi_{\boldsymbol{\theta_A}} $ and clear the replay buffer \cite{PPO}
            \EndIf
        \EndFor
        \renewcommand{\algorithmicrequire} {\textbf{Output:}} 
        \Require The updated weights of the actor network $\pi_{\boldsymbol{\theta}_A}$.   
        \end{algorithmic}  
        \label{algorithmcompleteTraining}
    \end{algorithm}

\section{Local Resources Allocation} \label{local Lagrangian}
Given the selected user sets in one specific round $k$, local resource allocation can be exploited to further enhance the convergence performance. The corresponding local problem can be described as jointly optimizing the CPUs frequency, subcarrier assignment and transmission power of each users in the $k$-th round with the objective to minimize the duration of this $k$-th round time, which can be formulated as follows:
\begin{align}\label{one round initial problem}
    \textbf{P2}: \min_{f_{k,n}, c_{k,n}^m, P_{k,n}^m} \  &  \max_{n \in \mathcal{V}_k} 
    \left\{t_{k,n} ^\text{(cp)} +  t_{k,n} ^\text{(cm)} \right\} &\\
    \text{s.t.}\   \ &c_{k,n}^m \in \left\{ 0,1 \right\},\ \forall m, \ \forall {n \in \mathcal{V}_k} , &\tag{21a}\\
    & e_{k,n} ^\text{(cp)} + e_{k,n} ^\text{(cm)} \leq E_{k,n},\  \forall {n \in \mathcal{V}_k} , &\tag{21b}\\
    & \text{(\ref{initial problem}c) - (\ref{initial problem}f)}. & &\nonumber
\end{align}
 However, the optimization problem \textbf{P2} is hard to solve due to the non-convexity and binary nature of $c_{k,n}^m$. To address these complexities, the variable $c_{k,n}^m$ is relaxed to take real values within the interval $[0, 1]$. Additionally, a new auxiliary variable $u_{k,n}^m=P_{k,n}^mc_{k,n}^m$ is introduced to transform the problem into a convex one. Correspondingly, the achievable transmission rate can be written as $\Tilde{R}_{k,n} = \sum_{m=1}^{M} B c_{k,n}^m\log_2\left(1+\frac{u_{k,n}^m   \varphi_{k,n}^m }{c_{k,n}^m}\right)$.  Then, to smoothen the  objective function, the following proposition is introduced.
\newtheorem{proposition}{Proposition}
\begin{proposition} \label{proposition1}
    For the problem \textbf{P2}, the optimized total computation and communication time is equal for all scheduled users, i.e., there exists a fixed value $t_k^{*}$ such that $t_k^{*}=t_{k,n} ^\text{(cp)}+t_{k,n} ^\text{(cm)}, \ \forall n \in \mathcal{V}_k$.
\end{proposition}
\begin{IEEEproof}
    See Appendix \ref{appendixB}.
\end{IEEEproof} 
However, the challenge persists due to the interdependent constraints on computational and communication resources in the objective function and (\ref{one round initial problem}b). To address this coupling, the problem is decomposed into two sub-problems: CPUs frequency control and communication resources allocation. An Alternating Direction Optimization (ADO) method is then applied to iteratively optimize each set of variables while keeping the others constant until convergence, as outlined in \textbf{Algorithm~\ref{algorithmado}}. Notably, optimal solutions are attained in each sub-process, thereby ensuring convergence of the ADO algorithm \cite{yangzhaohui, yaojingjing}. Since the optimization is applicable to any $k$-th round, the subscript $k$ is omitted in the remaining sections when no ambiguity arises. 
      \begin{algorithm}[t]  
         \caption{Alternating Direction Optimization (ADO)}
        \begin{algorithmic}[1]  
        \Ensure initialize the iteration step index $\iota=0$ and a feasible solution set $f_{n}^{(0)}=\left( f_{n}^{\text{min}} + f_{n}^{\text{max}}\right)/2$.
        \While{algorithm has not converge}
            \State Calculate the optimal $(P_{n}^{m})^{(\iota)}$ and $(c_{n}^{m})^{(\iota)}$ according\Statex \ \ \ \,  to LDRA or LCRA
            \State Calculate the optimal $f_n^{(\iota+1)}$ based on (\ref{optimal f})
            \If {$\iota \ge \iota_\text{max}$}
            \State break
            \EndIf
            \State $\iota = \iota+1$
        \EndWhile
        \renewcommand{\algorithmicrequire} {\textbf{Output:}} 
        \Require local resource allocation solution.   
        \end{algorithmic}  
        \label{algorithmado}
    \end{algorithm} 
    
\subsection{CPUs Frequency Optimization}
To optimize the CPUs frequency, the problem can be formulated as:
\begin{align}\label{computation}
    \textbf{P2-1}: \min_{f_{n}}& \ \max_{n \in \mathcal{V}_k} \left\{ \frac{Ec_nd_n}{f_n}+\left(t_{n} ^\text{(cm)}\right)^* \right\}&\\
    \text{s.t.}\  &\ {\kappa E c_n d_n}{{f_n^2}}+\left(e_{n} ^\text{(cm)}\right)^*\leq E_{n},\ \forall {n \in \mathcal{V}_k}, &\tag{22a}\\
    &f_{\text{min}}\leq f_{n} \leq f_{\text{max}}, \hfill  \ \forall {n \in \mathcal{V}_k}.\hfill &\tag{22b} \nonumber
\end{align}
Notably, the constraints can be combined as the following:
\begin{equation} \label{optimal cpu}
    f_{\text{min}}\leq f_{n} \leq \min \left\{ \sqrt{\frac{E_{n}-\big(e_{n} ^\text{(cm)}\big)^*}{\kappa E c_n d_n}}, f_{\text{max}}\right\}.
\end{equation}
Under \textit{Proposition~\ref{proposition1}}, the optimized solutions for \textbf{P2-1} can be given by:
\begin{equation}
    t^*=\max\left\{\frac{Ec_nd_n}{{f_n^{\text{max}}}}+\left(t_{n} ^\text{(cm)}\right)^*\right\},
\end{equation}
and
\begin{equation} \label{optimal f}
    f_n^* = \frac{E c_n d_n}{t^*-\left(t_{n} ^\text{(cm)}\right)^*},
\end{equation}
where ${f_n^{\text{max}}}$ can be calculated through (\ref{optimal cpu}).

\subsection{Communication Process Optimization}
To optimize the communication resource allocation process, the following subproblem should be considered:
\begin{align}\label{local communication resource allocation}
       \textbf{P2-2}: \min_{c_{n}^m, P_{n}^m}\quad &\max_{n \in \mathcal{V}_k} \left\{ \left(t_{n} ^\text{(cp)}\right)^*  + \frac{\Pi}{\Tilde{R}_{n}}\right\} & &\\
        \text{s.t.} \quad 
        & \left(e_{n} ^\text{(cp)}\right)^*+e_{n} ^{\text{(cm)}}  \leq E_{k,n},\  \forall {n \in \mathcal{V}_k},  &\tag{26a}\\
        & \ \text{(\ref{initial problem}c) - (\ref{initial problem}e)}. \nonumber
\end{align}
Then, considering \textit{Proposition~\ref{proposition1}}, a new variable  $\varsigma>0$ is introduced, such that \textbf{P2-2} can be smoothly recast as:
    \begin{align} \label{p2-2}
         \min_{c_{n}^m,P_{n}^m}  &\quad\ \  \varsigma \\
        \text{s.t.} \ & \quad  \frac{\Pi}{\varsigma-\left(t_{n} ^\text{(cp)}\right)^*}-\Tilde{R}_{n} \leq 0, \ \forall {n \in \mathcal{V}_k},& \tag{27a}\\
            &\quad\sum\nolimits_{m=1}^{M} P_{n}^mc_{n}^m \Pi \!-\! E_{n}^{\text{(cm)}} R_{n} \!\leq \!0, \ \forall {n \in \mathcal{V}_k}, \tag{27b}\\
            &\quad\text{(\ref{initial problem}c) - (\ref{initial problem}e)}, &\nonumber
    \end{align}
where $E_{n}^{\text{(cm)}}=E_{k,n} - \left(e_{n} ^{\text{(cp)}}\right)^*$ denotes the energy left for communication. 
It should be noted that the power allocation and subcarrier assignment are coupled in constraints (\ref{p2-2}a) and (\ref{p2-2}b), making the combinational optimization of these two variables challenging. Hereinafter, two methods named LDRA and LCRA approach are proposed separately to solve this.
\subsubsection{Lagrangian Decomposition-based Resource Allocation}
A conventional LDRA approach which iteratively obtains a suboptimal solution with convergence guarantee \cite{bertsekas1997nonlinear} is proposed to tackle this problem. Since this method has been adopted in previous literature \cite{793310}, though with different scenarios and motivations, the detailed procedure is presented in Appendix~\ref{appendixC} with the key steps listed in \textbf{Algorithm~\ref{algorithmlagrangian}}.
       \begin{algorithm}[t]  
        \caption{Lagrangian Decomposition-based Resource Allocation (LDRA)}
        \begin{algorithmic}[1]  
        \Ensure initialize the Lagrangian multiplier and set the iteration step index $\iota=0$.
        \While{algorithm has not converge}
            \State For each subcarrier, calculate the optimal power \Statex \ \ \ \, allocation in (\ref{optimal p}) with the given Lagrangian \Statex \ \ \ \, multipliers
            \State Calculate the subcarrier allocation matrix $V_{k,n}^m$ \Statex \ \ \ \, following (\ref{optimal h}) and allocate the subcarrier according \Statex \ \ \ \, to (\ref{optimal c})
            \State For each user, update the Lagrangian multipliers $\lambda_n(\iota)$,\Statex \ \ \ \, $\gamma_n(\iota)$, $\mu_n(\iota)$ according to  (\ref{update lamda}) - (\ref{update mu}), respectively
            \If {$\iota \ge \iota_\text{max}$}
            \State break
            \EndIf
            \State $\iota = \iota+1$
        \EndWhile
        \renewcommand{\algorithmicrequire} {\textbf{Output:}} 
        \Require Subcarrier and power allocation solution.   
        \end{algorithmic}  
        \label{algorithmlagrangian}
    \end{algorithm} 

\subsubsection{Low Complexity Resource Allocation}
While the proposed LDRA demonstrates feasibility, its convergence requires multiple iterations, potentially causing delays in model transmission. To improve the algorithm's suitability for time-sensitive tasks, a suboptimal LCRA method is introduced. Specifically, the LCRA consists of two separate processes: 
\textit{i)} initial subcarrier and power assignment and \textit{ii)} power adjustment. The detailed process is described in the following:
 
    \textit{i)}\  \textit{Initial Subcarrier and Power Assignment:} This step focuses on the allocation of individual subcarrier and the corresponding power to each user, with the primary goal of enhancing the communication stragglers. Users are sorted based on their descending pathloss values, and each user is sequentially allocated the best subcarrier that ensures the highest possible transmission rate for that user. Then, power is allocated according to the following proposition.
   \begin{proposition} \label{proposition2} 
  The optimal power allocation of LDRA in (\ref{optimal p}) indicates that the power allocation for user $n$ can be achieved via water-filling algorithm. Specifically, the water line for all the allocated subcarriers $\mathcal{U}_n$ for user $n$ is fixed and can be given as:
\begin{equation} \label{lowcomplexity power allocation}
  \theta_n^\text{WL} = \min \left\{ \theta_n^{\text{E}}, \theta_n^{\text{P}} \right\},
\end{equation}
\end{proposition}
where $\theta_n^\text{E}$ and $\theta_n^\text{P}$ represent the water levels determined by the energy constraint (\ref{local communication resource allocation}a) and power constraint (\ref{initial problem}c), respectively.
\begin{IEEEproof}
    See Appendix \ref{appendixD}.
\end{IEEEproof}
In this stage, every user is guaranteed at least one subcarrier for transmission (line 1-4 in \textbf{Algorithm~\ref{algorithmlowcomplexity}}). After the initial allocation, the user with the lowest transmission rate is identified. The BS then selects a subcarrier from the available pool that can provide the maximum transmission rate for this user. The transmission power for this user on each subcarrier will be updated according to \textit{Proposition~\ref{proposition2}}. This process continues iteratively until all available subcarriers are allocated to the users or can no longer be allocated due to users' energy and maximum power limitations. After this process, each user will have a set of allocated subcarriers $\mathcal{U}_n$ and their corresponding maximum achievable transmission rate $R_n^\text{max}$.

\textit{ii)}\  \textit{Power Adjustment:}
Once the subcarriers are allocated to the users and the global transmission time is determined, the next step is to lower the transmission power on the assigned subcarriers for each user to minimize energy consumption. This is because the global optimal time is determined by the worst computation and communication straggler, whereas the other users will simply be idle once their tasks are completed. Therefore, these other users can reduce their transmission power such that they also finish at the same time. Specifically, the global optimal time due to the worst straggler is given as:
\begin{equation} \label{lowcomplexity time}
    t^*_\text{LCRA} = \max_{n \in \mathcal{V}_k} \left\{ \left(t_{n} ^\text{(cp)}\right)^*  + \frac{\Pi}{{R}_{n}^\text{max}} \right\}.
\end{equation}
Then, the water line for each user can be lowered to match the global optimal time $t^*$, which is given as:

\begin{equation} \label{low complexity synchronize power}
    \theta_n^\text{syn} = 2^{\frac{A_n}{B_n}},
\end{equation}
where $A_n = \Pi - t^*_\text{LCRA}\sum_{m \in \mathcal{U}_n}B\log_2(\varphi_n^{m})$ and $B_{n_l}=|\mathcal{U}_n|Bt^*_\text{LCRA}$. 
The detailed LCRA is summarized in \textbf{Algorithm~\ref{algorithmlowcomplexity}}. The allocated power for the assigned subcarriers for user $n$ can then be updated as $\theta_n^\text{syn}-1/\varphi_n^{m},\  \forall m \in \mathcal{U}_n$.

\begin{algorithm}[t]  
    \caption{Low Complexity Resource Allocation (LCRA)}
    \begin{algorithmic}[1]  
    \Ensure initialize an empty subcarrier list $\mathcal{U}_n$ for each user. Sort users based  in descending order of pathloss values and create a queue $\Upsilon$.
    \Statex \textit{\textbf{\ \ \ i): Initial Subcarrier and Power Assignment}}
    \For{each user $n$ in $\Upsilon$}
    \State Allocate one subcarrier that maximizes the transmis-  \Statex \ \ \ \,  sion rate for user $n$
    \State Add the allocated subcarrier to the user's allocated su-\Statex \ \ \ \,  bcarrier list $\mathcal{U}_n$ and remove the subcarriers  from the \Statex \ \ \ \,  global available pool
    \State Perform power allocation based on (\ref{lowcomplexity power allocation})
    \EndFor
\While{there are still unallocated subcarrier \textbf{and} incomplete \Statex  users}
    \State Identify the user with the lowest transmission rate: \Statex \quad \, $n_l = \arg\min\limits_{n}R_{n}$
    \State Select a subcarrier $m$ from the global available pool \Statex\quad  \, that offers the maximum channel gain for user $n_l$
    \If{$\theta_{n_l}^\text{WL}>\frac{1}{\varphi_{n_l}^m}$}
        \State Add subcarrier $m$ to $\mathcal{U}_{n_l}$
        \State Update the power allocation for the updated sub- \Statex \qquad \ \ \ carrier set $\mathcal{U}_{n_l}$ based on (\ref{lowcomplexity power allocation})
        \Else
        \State Mark the subcarrier allocation for $n_l$ as completed
    \EndIf
\EndWhile
    \Statex \textit{\textbf{\qquad \qquad \ \ \ \ \ ii): Power Adjustment}}
    \State Calculate the global optimal time through (\ref{lowcomplexity time})
    \State Synchronize the time for each user and update the new \Statex  water line and power allocation according to (\ref{low complexity synchronize power})
    \renewcommand{\algorithmicrequire}{\textbf{Output:}} 
    \Require Final subcarrier and power allocation solution.   
    \end{algorithmic}  
    \label{algorithmlowcomplexity}
\end{algorithm} 

\section{Simulation Results} \label{simulation results}
In this section, the effectiveness of the proposed adaptive biased user scheduling and resource allocation scheme is evaluated on the MNIST and CIFAR-10 databases. The to be trained FL model architecture comprises two convolutional layers, two pooling layers, and two fully connected layers. The dataset is distributed among  $N=20$ wireless users, with each user holding a subset of the data characterized by Non-IID data distribution. To introduce Non-IIDness, an allocation indicator $a$ is employed \cite{zhangjie}, wherein each user's dataset, containing $D_n$ that uniformly selected from $[200,500]$ samples, involves a dominant digit which number is determined by $a D_n$, with the remaining digits constituting the remainder. The training process takes into account the CPUs frequency of each user, which varies within $[\, 0.5,\, 3\,]$ GHz, with an effective switched capacitance of $\kappa=10^{-28}$. Additionally, the number of local epoch $E$ for one global round is 8, and the parameter $C_n$ is set to $20$ cycles$/$bit \cite{zhanyufengl4l}. 
For the channel model, the equation $38.4 + 30 \log_{10}(d_k) + X_\sigma$ is used to model the large scale fading, where $X_\sigma$ is the log-normal shadowing effect with zero mean and standard deviation 6 dB, and Rayleigh fading which has zero mean and unit variance is used to model the small scale fading. The users' maximum transmission power is $P_n^\text{max}=1$W, initial energy budget $E_{0,n}$ is randomly selected from $[0.5,\,1]$J with the expected EH value of $\varkappa=0.2$, and $N_0 = -174$ dBm$/$Hz, There is a total of $64$ subcarriers and the bandwidth of one subcarrier is $15$kHz. The uploaded model size is $\Pi=51.2$ kbits.
Regarding the training of DRL agent, the number of nodes of the actor and critic's hidden layer is 64, the learning rate is $3\times10^{-4}$, the mini-batch
size is 64 and the replay buffer size is 1000. 
In order to limit the output within $[0,\,1]$, the action will be sampled according to the beta distribution \cite{chou2017improving}. 

The proposed algorithm is compared with the following benchmarks that stems from  state-of-the-art algorithms:
\begin{itemize}
    \item Greedy \cite{Nishio}: Greedily schedule the maximum number of users with a given time threshold of $3$s.
    \item FedAvg \cite{zhanghangjia}: Schedule a fixed percentage of users (10 users) in each round to minimize single round time.
    \item Max Gradient \cite{chenmingzheconvergencetime,9904868}: Schedule a certain proportion of users (10 users) with the probabilities proportional to their gradient norm.
    \item Ascend \cite{xujie}: Schedule an ascending number of users with an average of 10 users.
    \item Unbiased Adaptive \cite{10443546}: Adaptive unbiased user scheduling with jointly heterogeneous consideration.
    \item Dynamic \cite{yinbenshun}: Dynamically scheduling a certain portion of users (10 users) based on the DRL scheduler.
\end{itemize}

\begin{figure}[htbp!]
      \centering
      \includegraphics[width=8.6cm]{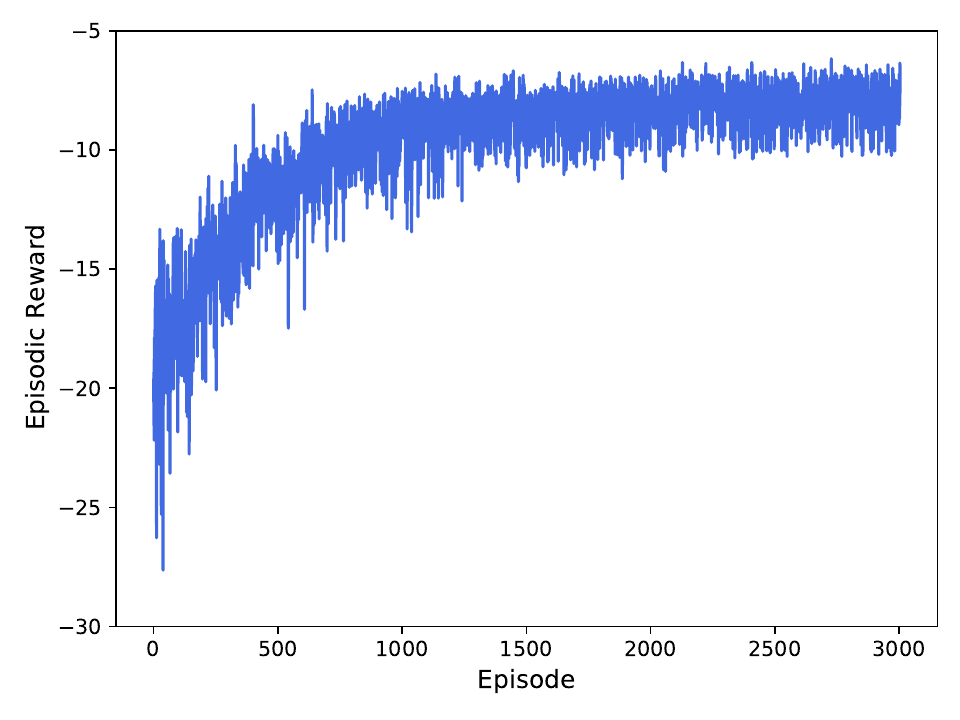}
      \caption{Training Convergence of the PPO-based DRL agent.}
      \label{fig: reward_fading}
\end{figure}

In the first simulation, the convergence performance of the PPO algorithm on MNIST dataset is validated, wherein the target accuracy $0.92$, the Non-IID ratio $a=0.8$ and each DRL training episode corresponds to one complete FL task.
As shown in Fig.~\ref{fig: reward_fading}, the rewards increase gradually, achieving stable rewards with slight fluctuations after approximately 1200 episodes. This indicates that the DRL agent is learning and adopting its policy to minimize the convergence wall-clock time of a FL task. The adoption of mini-batch SGD for training FL local models introduces random noise, contributing to fluctuations in FL accuracy. Consequently, the FL trainer may require one or more rounds to attain the target accuracy, resulting in reward variability. Additionally, the presence of a fading channel introduces varied transmission delays during each global round, further enriching the dynamic challenges faced by the agent. Despite these complexities, the DRL agent showcases its adaptability and robustness, ensuring the maintenance of accelerated convergence wall-clock time in FL tasks. 

\begin{figure}[tb]
    \centering
    \subfigure[]{
    \begin{minipage}{4cm}
    \centering
    \includegraphics[width=4cm]{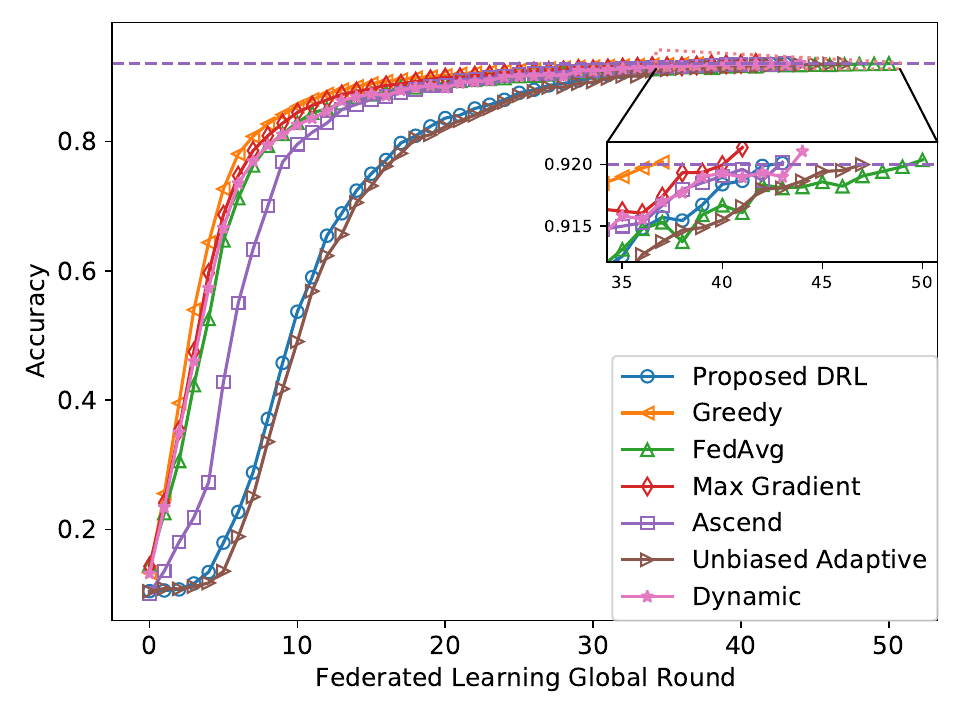}
    \end{minipage}%
    }%
    \subfigure[]{
    \begin{minipage}{4cm}
    \centering
    \includegraphics[width=4cm]{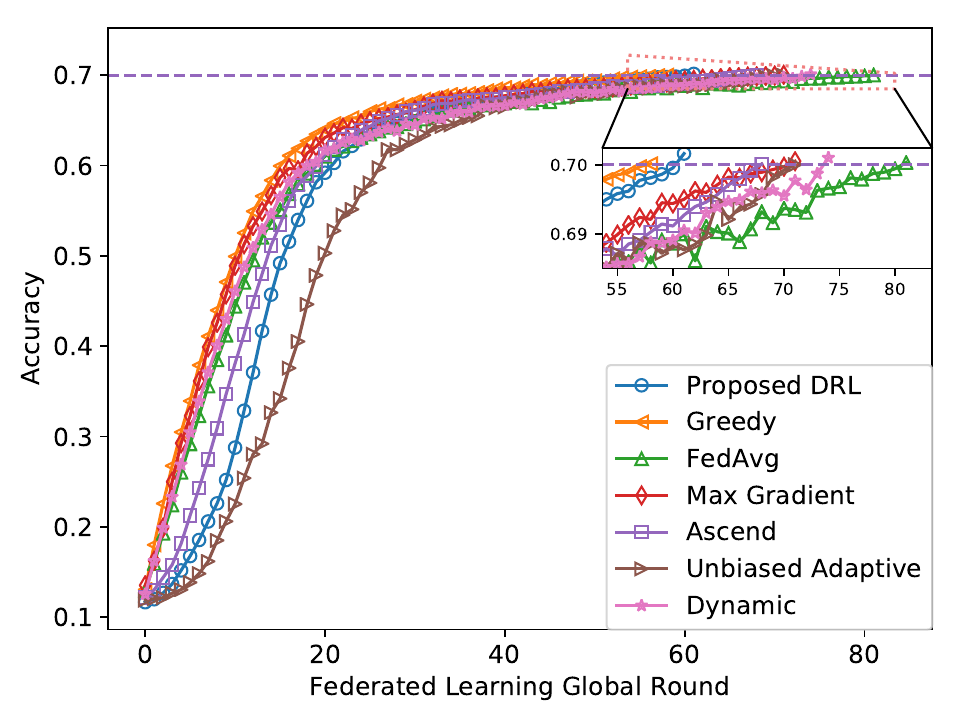}
    \end{minipage}
    }
    \caption{FL convergence performance on (a) MNIST and (b) CIFAR-10.}
    \label{fig: 2}
    \end{figure}

Then, the well-trained actor network undergoes testing across 100 different FL tasks characterized by the Non-IID data ratio $a$ of 0.8, with dynamic communication channels and randomized data distribution among users for each task. The FL convergence performance of the proposed DRL algorithm is evaluated in Fig.~\ref{fig: 2}  and compared with the benchmarks on MNIST and CIFAR-10, respectively. Notably, the proposed DRL scheme achieves target accuracy within 42 rounds on MNIST, matching the performance of the Ascend benchmark, while trailing the Max Gradient and Greedy benchmarks by 2 rounds and 6 rounds, respectively. On the CIFAR-10 dataset, the proposed scheme reached a target accuracy (70$\%$) within 61 rounds, outperforming all benchmark schemes except Greedy.

\begin{figure}[tb]
    \centering
    \subfigure[]{
    \begin{minipage}{4cm}
    \centering
    \includegraphics[width=4cm]{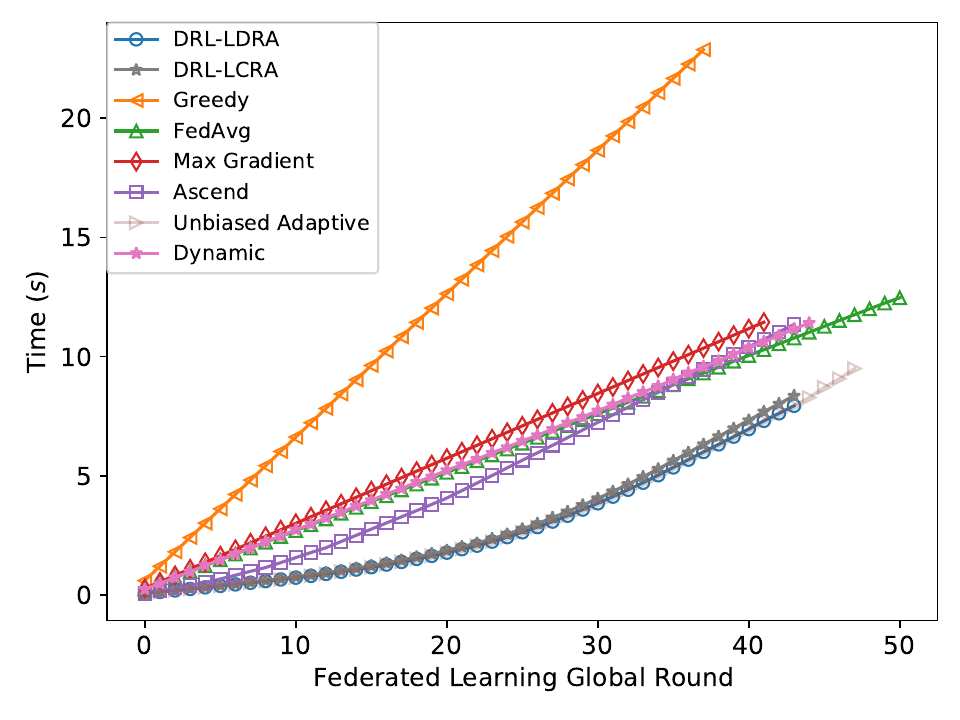}
    \end{minipage}%
    }%
    \subfigure[]{
    \begin{minipage}{4cm}
    \centering
    \includegraphics[width=4cm]{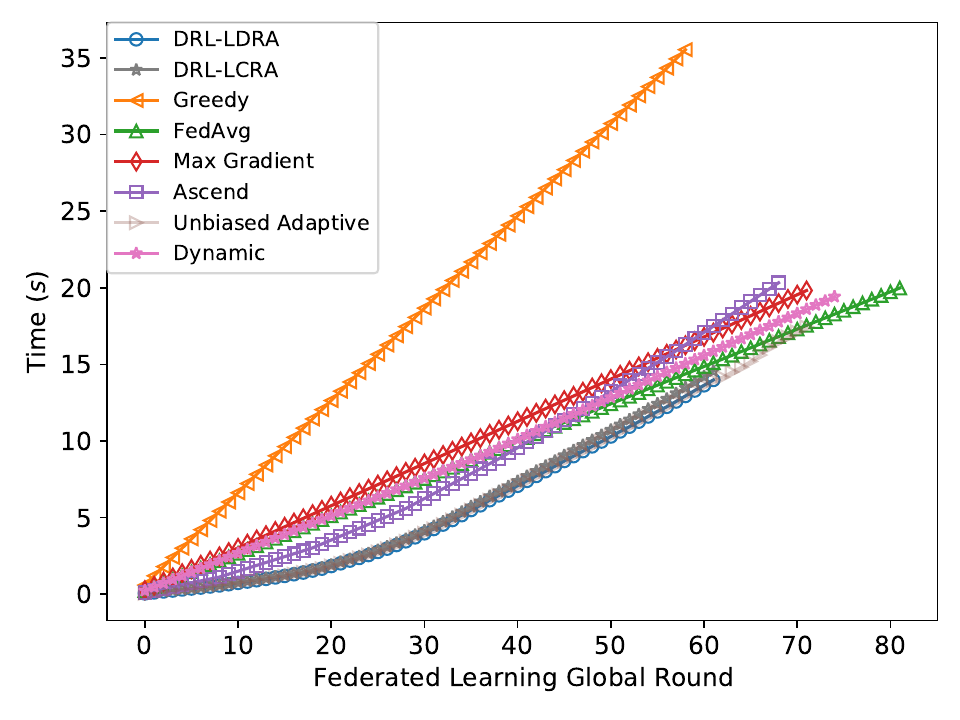}
    \end{minipage}
    }
    \caption{FL convergence time verses the iteration round of one FL task on (a) MNIST and (b) CIFAR-10.}
    \label{fig: 3}
\end{figure}

While the trained DRL algorithm lags behind in the total number of rounds compared with  Max Gradient and Greedy benchmarks on MNIST and compared with Greedy on CIFAR-10, it excels at striking the balance between the single round time and the overall global rounds to achieve a faster convergence time. This is evident in Fig.~\ref{fig: 3}, where the proposed DRL-LCRA and DRL-LDRA outperform other benchmarks in terms of the total run time for a single FL task in both datasets. This assessment underscores the effectiveness of the proposed scheme, as it reduces the total run time by more than $40\%$ compared to the Max Gradient and Ascend benchmarks, and by nearly $200\%$ compared to the Greedy scheme on MNIST. Notably, DRL-LCRA achieves similar performance to DRL-LDRA, but significantly reducing the algorithm computation load on the BS. 
Similar results can be observed in CIFAR-10 dataset, which validate the effectiveness of the proposed algorithm in different FL tasks.

\begin{figure}[tb]
    \centering
    \subfigure[]{
    \begin{minipage}{4cm}
    \centering
    \includegraphics[width=4cm]{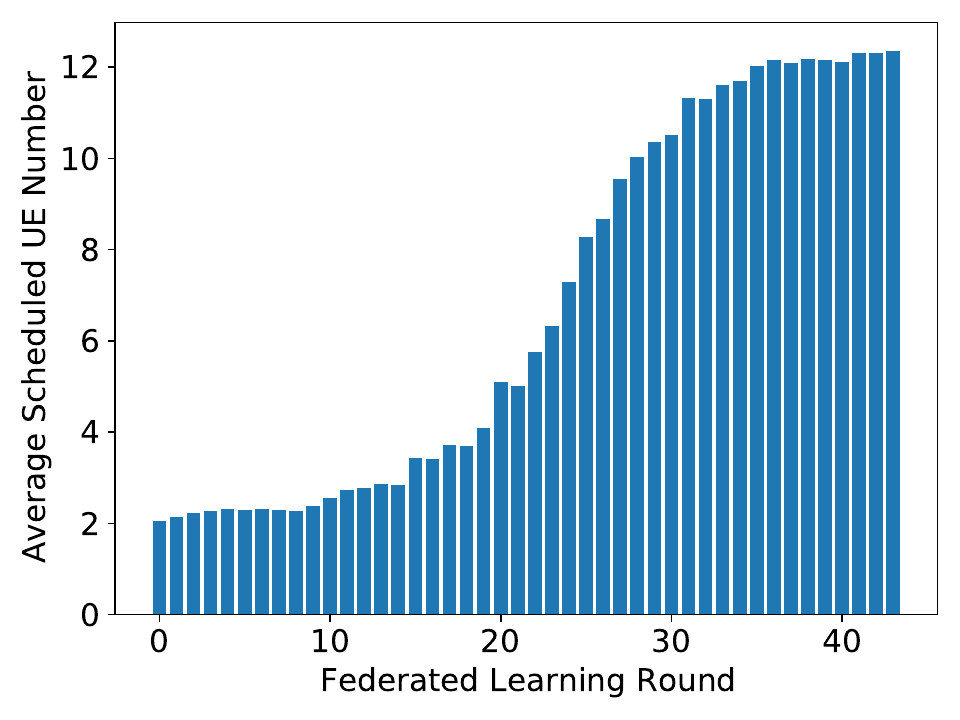}
    \end{minipage}%
    }%
    \subfigure[]{
    \begin{minipage}{4cm}
    \centering
    \includegraphics[width=4cm]{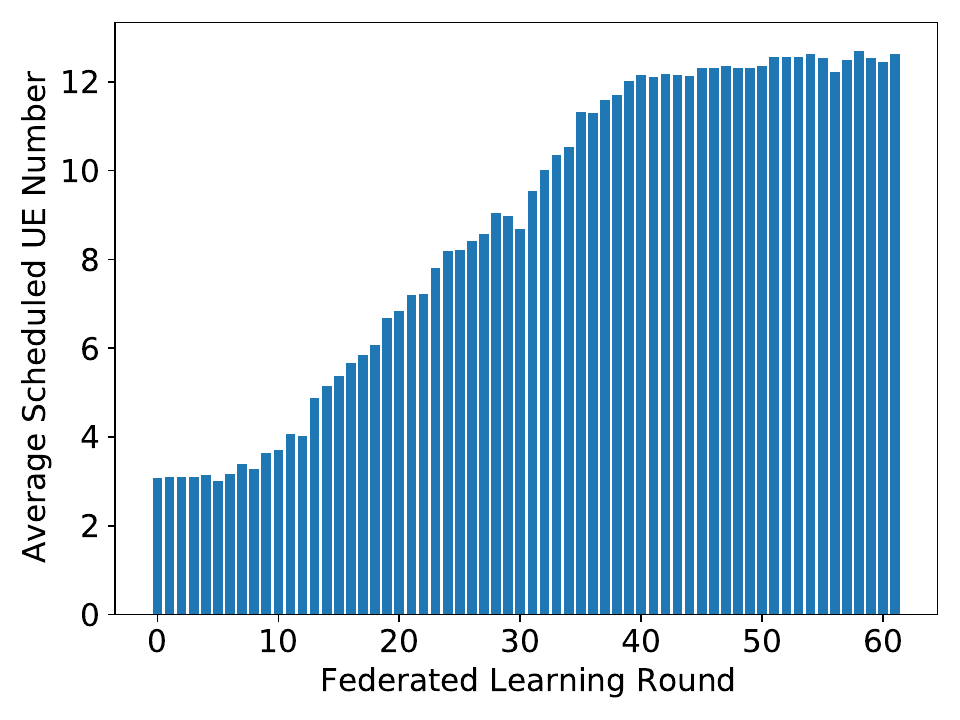}
    \end{minipage}
    }
    \caption{Scheduled user number verses the iteration round of one FL task on (a) MNIST and (b) CIFAR-10.}
    \label{fig: 4}
\end{figure}

To gain insight into the rationale behind the intelligent scheduling scheme, Fig.~\ref{fig: 4} illustrates the scheduled number of users in each global round. It is evident that at the beginning, only less than 4 users are selected for both datasets. This initial selection strategy is aimed at mitigating the influence of stragglers because the bias introduced by absent users has less influence on the final result. Subsequently, more users are included in the middle rounds to expedite the training process, ultimately reducing the total required rounds. In summary, the DRL-based algorithm effectively leverages straggler to strike a balance between individual round duration and the overall number of rounds, leading to a significant reduction in the overall time needed for convergence.

\begin{figure}[tb]
    \centering
    \subfigure[]{
    \begin{minipage}{4cm}
    \centering
    \includegraphics[width=4cm]{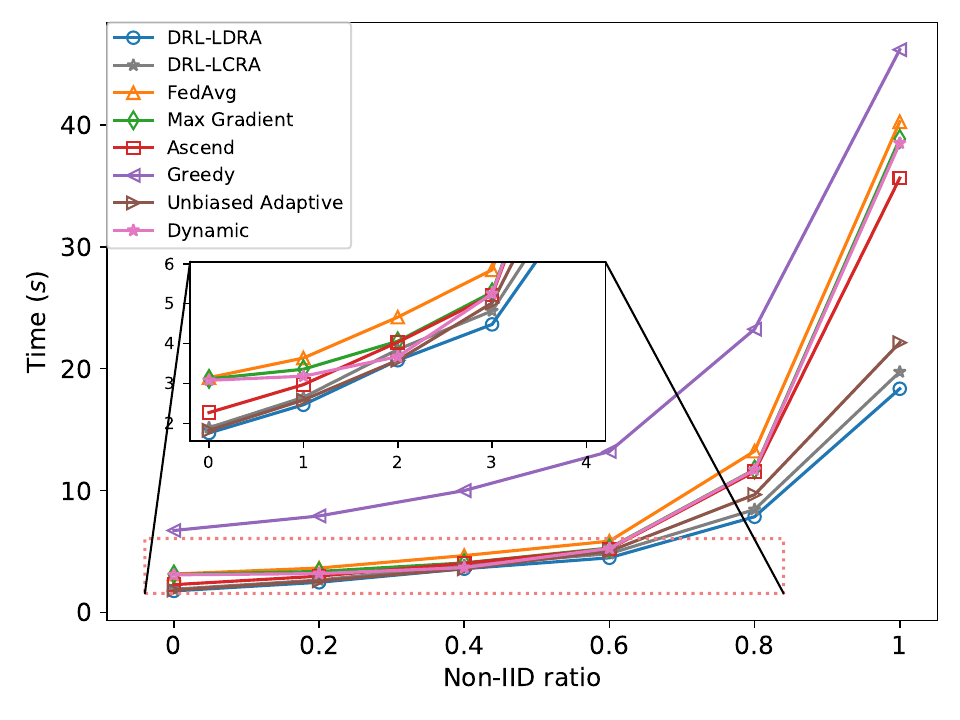}
    \end{minipage}%
    }%
    \subfigure[]{
    \begin{minipage}{4cm}
    \centering
    \includegraphics[width=4cm]{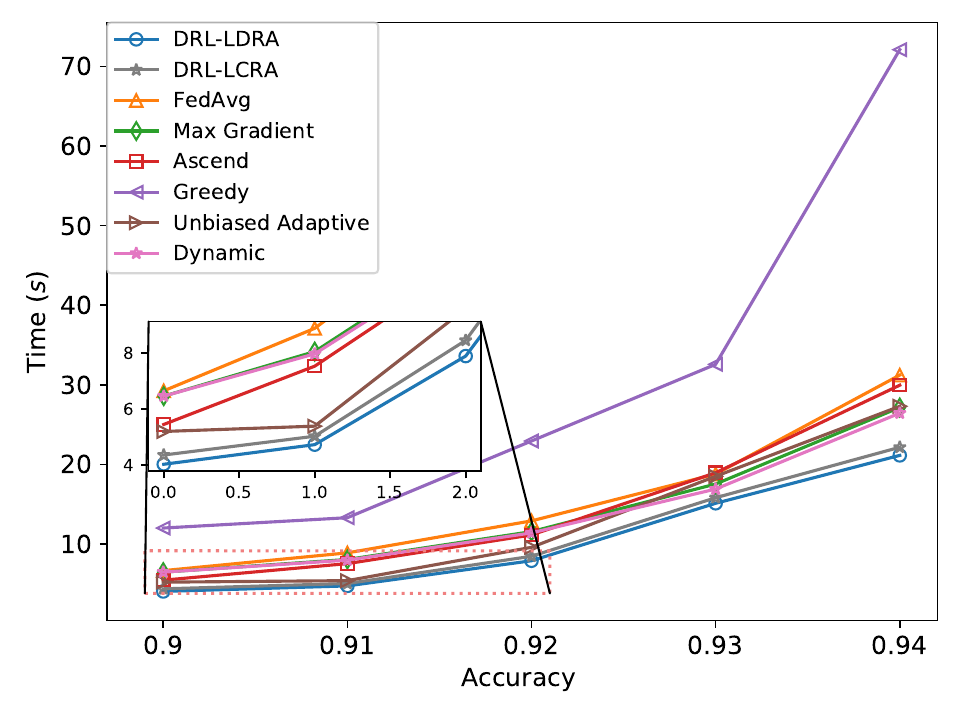}
    \end{minipage}
    }
    \caption{Convergence time versus (a) Non-IID ratio and (b) target accuracy.}
    \label{fig: 5}
\end{figure}

In the final evaluation, the proposed algorithm undergoes testing across various FL tasks characterized by diverse Non-IID data distributions and target accuracy. Fig.~\ref{fig: 5} illustrates that the proposed DRL-LDRA and DRL-LCRA consistently outperform the benchmarks across different Non-IID ratios and accuracy. This consistent superiority underscores the robustness of the proposed algorithm, demonstrating its ability to adapt effectively to varying data or targets scenarios.

\section{Conclusion} \label{conclusion}
This paper investigates the challenge of handling stragglers to expedite convergence process in wireless FL networks characterized by both system and statistical heterogeneity. The objective is to minimize convergence latency by orchestrating a comprehensive approach that encompasses user scheduling, subcarrier assignment, CPUs frequency, and transmission power allocation strategies. To cope with the ever-evolving dynamic updates and uploads scenario, a novel DRL-based user scheduling scheme is introduced, with the goal to reduce the long-term computation and communication time consumption. To bolster the robustness of the DRL agent, the PPO approach is employed for the agent training, resulting in an autonomous policy maker capable of navigating dynamic scenarios and online scheduling. Following user scheduling, two resource allocation strategies are presented: a traditional LDRA and a decoupled LCRA. These strategies can further enhance computation and communication efficiency while LCRA enables the reducing computation complexity on the BS. The extensive simulation results demonstrate the effectiveness and robustness of our proposed scheme across various Non-IID settings, showcasing remarkable time reductions ranging from $40\%$ to $230\%$.

\appendices 
\section{} \label{appendixA}
\subsection{Simplified Notations}
Before conducting the convergence analysis, the following short notations are introduced to simplify the expressions:

${\Tilde{g}}_{k}^{} = \frac{1}{V_k} \sum_{n\in\mathcal{V}_k} {\Tilde{g}}_{k}^{n} = \frac{1}{V_k} \sum_{n\in\mathcal{V}_k} \nabla{l_{\mathcal{B}_{k,n}}(\boldsymbol{w}_{k,n}^{})}$ represents the average stochastic gradients over the selected users in the $k$-th round.

${g}_{k}^{} = \mathbb{E}_{\mathcal{B}}[{\Tilde{g}}_{k}^{}] = \frac{1}{V_k} \sum_{n\in\mathcal{V}_k} {g}_{k}^{n} = \frac{1}{V_k} \sum_{n\in\mathcal{V}_k} \nabla{F_{n}(\boldsymbol{w}_{k,n}^{})}$ represents the average gradients over the selected users in the $k$-th round.

For the purpose of convergence analysis, a virtual global model $\Bar{\boldsymbol{w}}$ is introduced, which is assumed to be aggregated at an arbitrary time slot $k$. Based on (\ref{localupdate}) and (\ref{aggregation}), along with the above simplified notations, the progression of the global model between two consecutive time slots can be expressed as:
\begin{equation}
    \Bar{\boldsymbol{w}}_{k+1} = \Bar{\boldsymbol{w}}_{k} -  \frac{\eta_k}{V_k}\! \sum_{n\in\mathcal{V}_k} \!   {\Tilde{g}}_{k}^{n}.
\end{equation}

\subsection{Proof of Theorem 1}
\begin{IEEEproof}
Using the second-order Taylor expansion and \textbf{Assumption}~\ref{assumption1}, expanding the loss function around $\boldsymbol{w}_k$, we obtain:
        \begin{align} \label{taylor}
        &F(\Bar{\boldsymbol{w}}_{k+1}^{})-F(\Bar{\boldsymbol{w}}_{k}^{})\nonumber\\
        \leq& \left\langle \nabla F (\Bar{\boldsymbol{w}}_{k}^{}),\ \Bar{\boldsymbol{w}}_{k+1}^{}-\Bar{\boldsymbol{w}}_{k}^{}\right\rangle +\frac{\beta_k}{2} \left\| \Bar{\boldsymbol{w}}_{k+1}^{}-\Bar{\boldsymbol{w}}_{k}^{}\right\|^2\nonumber\\
        =&\left\langle \nabla F(\Bar{\boldsymbol{w}}_{k}),- \frac{\eta_k}{V_k}\!\sum_{n\in\mathcal{V}_k} \!   {\Tilde{g}}_{k}^{n} \right\rangle \!+\! \frac{\beta_k}{2}\left\|   \frac{\eta_k}{V_k}\! \sum_{n\in\mathcal{V}_k} \!  {\Tilde{g}}_{k}^{n} \right\|^2.
        \end{align}  
    Leveraging the fact that $(a-b)^2=a^2+b^2-2ab$ and taking  the expectation, the first term on the Right Hand Side (RHS) of (\ref{taylor}) can be rewritten as:
    \begin{align} \label{1term}
        &-\eta_k\left\langle \nabla F(\Bar{\boldsymbol{w}}_{k}), \mathbb{E}\left[\frac{1}{V_k}\sum_{n\in\mathcal{V}_k} \!   {\Tilde{g}}_{k}^{n}\right] \right\rangle=\frac{\eta_k}{2}\Bigg( \Bigg\|  \sum_{n=1}^N    p_{k}^{n} g_k^n -\!   \nonumber\\
        &\ \ \nabla F(\Bar{\boldsymbol{w}}_{k}) \Bigg\|^2-\!\left\|   \sum_{n=1}^N    p_{k}^{n} g_k^n \right\|^2 \!-\! \left\| \nabla F(\Bar{\boldsymbol{w}}_{k}) \right\|^2 \Bigg).
    \end{align}
    The first term on the RHS of (\ref{1term}) can be bounded as follows:
    \begin{align} \label{convergence temp1}
        &\left\|  \sum_{n=1}^N\! p_k^n   {{g}}_{k}^{n} \!-\!  \nabla F(\Bar{\boldsymbol{w}}_{k}) \right\|^2 \nonumber\\
        =&\left\| \sum_{n=1}^{N} q_n g_{k}^{n}+ \sum_{n=1}^{N}(p_k^n-q_n)g_{k}^{n} -\nabla F \left( \Bar{\boldsymbol{w}}_{k} \right) \right\|^2  \nonumber\\
        \leq& \left( \left\| \sum_{n=1}^{N} q_ng_{k}^{n}-\nabla F \left( \Bar{\boldsymbol{w}}_{k} \right) \right\| + \sum_{n=1}^{N}|p_k^n-q_n|\left\|g_{k}^{n}\right\|\right)^2 \nonumber\\
        =& \left( \sum_{n=1}^{N} q_n\left\| g_{k}^{n}-\nabla F \left( \Bar{\boldsymbol{w}}_{k} \right) \right\| + \sum_{n=1}^{N}|p_k^n-q_n|\left\|g_{k}^{n}\right\|\right)^2 \nonumber\\
        \leq&\left(\!\ 1+\sum_{n=1}^N(p_k^n-q_n)^2\right)\Bigg( \underbrace{\sum_{n=1}^{N} q_n\left\| g_{k}^{n}-\nabla F \left( \Bar{\boldsymbol{w}}_{k} \right) \right\|^2 }_{\text{(a)}}\nonumber\\
        &\ + \sum_{n=1}^N \| g_k^n\|^2\Bigg),
    \end{align}
    where the first inequality follows from the triangle inequality, while the second arises from the Cauchy–Schwarz inequality. The second equality holds as the user-set is scheduled independently across all users. Subsequently, the unbiased term (a) in (\ref{convergence temp1}) can be further expanded as:
    \begin{align} \label{convergence 3term}
        &\sum_{n=1}^{N} q_n\left\| g_{k}^{n}-\nabla F \left( \Bar{\boldsymbol{w}}_{k} \right) \right\|^2\nonumber\\
        =&\sum_{n=1}^{N} \frac{p_k^n}{\rho}\left\| g_{k}^{n} - \nabla F_n \left( \Bar{\boldsymbol{w}}_{k}\right) + \nabla F_n \left( \Bar{\boldsymbol{w}}_{k}\right) - \nabla F \left( \Bar{\boldsymbol{w}}_{k} \right) \right\|^2\nonumber\\
        \leq&\ \frac{2}{\rho}\sum_{n=1}^{N} p_k^n \left\| g_{k}^{n} - \nabla F_n \left( \Bar{\boldsymbol{w}}_{k}\right) \right\|^2+ \frac{2}{\rho N}\sum_{n=1}^{N} p_k^n\sum_{n=1}^{N}\nonumber\\&\  \left\| \nabla F_n ( \Bar{\boldsymbol{w}}_{k}\right) - \nabla F \left( \Bar{\boldsymbol{w}}_{k} \right) \|^2\nonumber\\
        \leq&\ \frac{2}{\rho}{\sum_{n=1}^{N} p_k^n \left\| g_{k}^{n} - \nabla F_n \left( \Bar{\boldsymbol{w}}_{k}\right) \right\|^2} + \frac{2}{\rho}\alpha \nonumber\\
        \leq&\ \frac{2\beta_k^2}{\rho}\underbrace{\sum_{n=1}^{N} p_k^n \left\| \boldsymbol{w}_{k}^{n} - \Bar{\boldsymbol{w}}_{k} \right\|^2}_{\text{(b)}} + \frac{2}{\rho}\alpha,
    \end{align}
    where the first equality  originates from \textbf{Definition~\ref{definition}}. The first inequality is derived using Jensen's inequality, the second follows from \textbf{Assumption~\ref{assumption3}}, and the third leverages \textbf{Assumption~\ref{assumption1}}. To further constrain the term (b) in the above inequality, it is assumed that model aggregation occurs at $\tau_0$, where $\tau_0 \leq k < \tau_0 + E$, and the derivation proceeds as follows:
    \begin{align} \label{convergence 4term}
        &\mathbb{E}\left[\sum_{n=1}^{N} p_k^n \left\| \boldsymbol{w}_{k}^{n} - \Bar{\boldsymbol{w}}_{k} \right\|^2\right]\nonumber\\
        =&\mathbb{E}\left[\sum_{n=1}^{N} p_k^n \left\| \boldsymbol{w}_{k}^{n} - \frac{1}{V_k}\sum_{n\in\mathcal{V}_k}\Bar{\boldsymbol{w}}_{k} \right\|^2\right]\nonumber\\
        =& \mathbb{E}\left[\sum_{n=1}^{N} p_k^n\left\|  \Bar{\boldsymbol{w}}_{\tau_0} \!-\! \eta_k\!\sum_{\tau=0}^{k-1} {\Tilde{g}}_{k,n}^{\tau} \!-\! \Bar{\boldsymbol{w}}_{\tau_0} \!+\! \frac{\eta_k}{V_k}\!\sum_{n\in\mathcal{V}_k} \sum_{\tau=0}^{k-1}\! {\Tilde{g}}_{k,n}^{\tau}  \right\|^2\right]\nonumber\\
        =&\mathbb{E}\left[\eta_k^2 \sum_{n=1}^{N} p_k^n\left\|   \sum_{\tau=0}^{k-1} {\Tilde{g}}_{k,n}^{\tau} - \frac{1}{V_k} \sum_{n\in\mathcal{V}_k} \sum_{\tau=0}^{k-1}\! {\Tilde{g}}_{k,n}^{\tau}  \right\|^2\right]\nonumber\\
        \leq& \mathbb{E} \left[2\eta_k^2  \Bigg( \sum_{n=1}^{N} p_k^n\left\| \sum_{\tau=0}^{k-1} {\Tilde{g}}_{k,n}^{\tau} \right\|^2 + \frac{1}{V_k}\sum_{n\in\mathcal{V}_k}\left\| \sum_{\tau=0}^{k-1} {\Tilde{g}}_{k,n}^{\tau} \right\|^2\Bigg)\right] \nonumber\\
        =&  4\eta_k^2 \sum_{n=1}^{N} p_k^n \mathbb{E}_\mathcal{B}\left\| \sum_{\tau=0}^{k-1} {\Tilde{g}}_{k,n}^{\tau} \right\|^2\nonumber\\
        =&  4 \eta_k^2 \sum_{n=1}^{N} p_k^n\mathbb{E}_\mathcal{B}\!\left( \sum_{\tau=0}^{E-1} \!\left\| {\Tilde{g}}_{k,n}^{\tau} - {g}_{k,n}^{\tau} \right\|^2 \!+\!  \sum_{\tau=0}^{E-1} \left\| {g}_{k,n}^{\tau} \right\|^2\!\right)\nonumber\\
        \leq& {4\eta_k^2 E \sigma} + 4{\eta_k^2}\sum_{n=1}^N p_k^n\sum_{\tau=0}^{E-1}\left\| {g}_{k,n}^{\tau} \right\|^2,
    \end{align}
   where the first inequality follows from Jensen's inequality, while the second stems from \textbf{Assumption~\ref{assumption2}}. The final equality holds due to the property $\mathbb{E}_{\mathcal{B}}[{\Tilde{g}}_{k,n}^{\tau}] - {{g}}_{k,n}^{\tau}=0$ and Jensen's inequality. Next, the expectation of the second term on the RHS of (\ref{taylor}) can be bounded as follows:
    \begin{align} \label{convergence 5term}
        &\mathbb{E}\left[\frac{\beta_k}{2}\left\|   \frac{\eta_k}{V_k}\! \sum_{n\in\mathcal{V}_k} \!  {\Tilde{g}}_{k}^{n} \right\|^2\right]\nonumber\\
        =\ & \frac{\beta_k \eta_k^2}{2 V_k^2} \mathbb{E}\left\| \sum_{n\in\mathcal{V}_k} \left( {\Tilde{g}}_{k}^{n} + {g}_{k}^{n} - {g}_{k}^{n} \right)\right \|^2 \nonumber\\
        =\ & \frac{\beta_k \eta_k^2}{2 V_k^2} \mathbb{E}\Bigg( \sum_{n\in\mathcal{V}_k}\left\|{\Tilde{g}}_{k}^{n} - {g}_{k}^{n} \right\|^2 + \sum_{n\in\mathcal{V}_k}\left\|{g}_{k}^{n} \right\|^2 \Bigg)
        \nonumber\\
        \leq\  & \frac{\beta_k \eta_k^2}{2V_k}\left(\sigma  +   \sum_{n=1}^N p_k^n \| g_k^n\|^2\right) \nonumber\\
        =\ & \frac{\beta_k \eta_k^2}{2V_k}\Bigg( \sigma + \left\| \sum_{n=1}^N p_k^ng_k^n\right\|^2 \!+\! \sum_{n=1}^N p_k^n\left\| g_k^n - \sum_{n=1}^N p_k^n g_k^n \right\|^2\!\Bigg),
    \end{align}
    where the last equality is because of the property $\mathbb{E}\left[x^2\right] = \mathbb{E}\left[x-\mathbb{E}[x]\right]^2 +\left(\mathbb{E}[x]\right)^2$.
    Assuming the learning rate $\eta_k < \frac{1}{\beta_k}$ and subsequently combining (\ref{1term}) - (\ref{convergence 5term}), while taking the expectation of (\ref{taylor}), we derive the bound on the loss function between two consecutive iterations as follows:
    \begin{align} \label{convergence one-round}
        &\mathbb{E}\left[F(\Bar{\boldsymbol{w}}_{k+1}^{})-F(\Bar{\boldsymbol{w}}_{k}^{})\right]\nonumber\\
        \leq & \frac{\eta_k}{2 V_k} \sigma + \frac{\eta_k}{2  V_k}\sum_{n=1}^N p_k^n\left\| g_k^n - \sum_{n=1}^N p_k^n g_k^n \right\|^2 \nonumber\\
        & + \frac{\eta_k}{2}\Bigg(\mathcal{E}\sum_{n=1}^N \| g_k^n\|^2 \!-\left\| \nabla F(\Bar{\boldsymbol{w}}_{k}) \right\|^2 \Bigg) \!+\! \frac{\eta_k\mathcal{E}}{2}\Bigg(\frac{2}{\rho}\Bigg({4 E \sigma}\nonumber\\
        & + 4\sum_{n=1}^N p_k^n\sum_{\tau=0}^{E-1}\left\| {g}_{k,n}^{\tau} \right\|^2\Bigg) + \frac{2}{\rho}\alpha\Bigg) \nonumber\\
        \leq&\! -\!\frac{\eta_\text{min}}{2}\left\| \nabla F(\Bar{\boldsymbol{w}}_{k}) \right\|^2 \!+\! \left( \frac{\eta_k}{2 V_k} + \frac{4 \eta_k E \mathcal{E}}{\rho} \right) \sigma \!+\! \frac{2 \eta_k }{V_k}G_k\nonumber \\
        & +\frac{\eta_k \mathcal{E} N}{2} {G_k} \!+\! \frac{4 \eta_k \mathcal{E}}{\rho}  \sum_{\tau=0}^{E-1}G_{k,\tau} \!+\! \frac{\eta_k \mathcal{E}}{\rho}\alpha,
    \end{align}
   where $G_k= \max_{n\in\mathcal{V}_k}$
   $\left\|  \nabla F_n({\boldsymbol{w}_k^n}) \right\|^2$ represents the gradient norm in the $k$-th round; $\eta_\text{min}$ is the minimum learning rate, and $\mathcal{E}=1+\sum_{n=1}^N(p_k^n - q_n)^2$. The last inequality is because of the decaying learning rate. Then summing up (\ref{convergence one-round}) while averaging over the total training rounds, the convergence bound in \textbf{Theorem~1} can be established.   
\end{IEEEproof}

\section{} \label{appendixB}
\begin{IEEEproof}
We prove \textit{Proposition~\ref{proposition1}} through a contradiction argument. On one hand, let's assume that there exists a user $n$ for which $t_{k,n}  < t_k^{*}, n\in\mathcal{V}_k$. In this case, decreasing the CPUs frequency or transmission power for user $n$ reduces their energy consumption without altering the optimal time, which contradicts the conditions for optimality. 
On the other hand, if $\Tilde{t}_{k,n}  > t_k^{*}$ for user $n$, then the optimal time should indeed be $\Tilde{t}_{k,n}$. This completes the proof.
\end{IEEEproof}

\section{} \label{appendixC}
\begin{IEEEproof}
 By introducing Lagrange multipliers, the constraints that associated with subcarrier and power allocation are relaxed, leading to the transformation of the original problem into two distinct problems: a suboptimal problem and a dual problem.
The Lagrangian function, derived from the optimization problem is given in (\ref{Lagrangian function}), where $\boldsymbol{\lambda}$, $\boldsymbol{\gamma}$, $\boldsymbol{\mu}$  and $\boldsymbol{\nu}$ are the Lagrange multipliers associated with the constraints on subcarrier and power allocation, respectively.
\begin{figure*}[ht] 
\centering 
\begin{align}
\label{Lagrangian function}
L(\boldsymbol{p},\boldsymbol{c},\boldsymbol{\lambda},\boldsymbol{\gamma},\boldsymbol{\mu},\boldsymbol{\nu}) =&  \varsigma + \sum_{n=1}^{N}\lambda_n\left(\frac{\Pi}{t-\left( t_n^{\text{(cp)}}\right)^*} - \Tilde{R}_{n}\right) +  
     \sum_{n=1}^{N}\gamma_n \left(\sum_{m=1}^{M}  u_{n}^m  \Pi -  E_{n}^{\text{(cm)}} \Tilde{R}_{n}\right) + 
     \sum_{n=1}^{N}\mu_n\left(\sum_{m=1}^{M}u_{n}^m- P_n^{\text{max}}\right)\nonumber \\ &+  \sum_{m=1}^{M}\nu_m\left(\sum_{n=1}^{N}c_{n}^m -1\right),
\end{align}
\vspace*{8pt} 
\hrulefill 
\end{figure*}
 The corresponding dual problem is:
\begin{equation}
    Z_D(\boldsymbol{\lambda},\boldsymbol{\gamma},\boldsymbol{\mu},\boldsymbol{\nu}) = \max L(\boldsymbol{p}^*,\boldsymbol{c}^*,\boldsymbol{\lambda},\boldsymbol{\gamma},\boldsymbol{\mu},\boldsymbol{\nu}).
\end{equation}
To solve the subproblem by differentiating the Lagrange function against $u_{k,n}^m$ and $c_{k,n}^m$ with a fixed set of Lagrangian multipliers, we have:
\begin{align} \label{s_equation}
    \frac{\partial L}{\partial u_{n}^m} &= -\frac{B  c_{n}^m  \varphi_{n}^m(\lambda_n+E_{n}^{\text{(cm)}} \gamma_n)}{\ln2 (c_{n}^m+u_{n}^m   \varphi_{n}^m)} + \mu_n + \Pi \gamma_n\nonumber\\
    &\begin{cases}
        > 0, \qquad  {\rm if} \ (u_{n}^m)^* = 0,\\
        = 0, \qquad  {\rm if} \ (u_{n}^m)^* \in (0, P_n^{\text{max}}),\\
        < 0, \qquad  {\rm if} \ (u_{n}^m)^* = P_n^{\text{max}},
    \end{cases}
\end{align}
and:
\begin{align} \label{c_equation}
    \frac{\partial L}{\partial c_{n}^m } &=     B(\lambda_n + \gamma_n E) \left[  \frac{u_{n}^m    \varphi_{n}^m }{\ln2(c_{n}^m +u_{n}^m    \varphi_{n}^m)} \nonumber\right.\\ 
    &\ \ \ \left.- \log_2\left(1+\frac{u_{n}^m    \varphi_{n}^m } {c_{n}^m }\right) \right] +\nu_m\nonumber\\
    &\begin{cases}
        = 0,   \qquad  {\rm if} \ (c_{n}^m)^* \in (0,1),\\
        < 0,  \qquad  {\rm if} \ (c_{n}^m)^* = 1.
     \end{cases}
\end{align}
Specifically, when $(c_{n}^m)^* = 0$,  $(u_{n}^m)^* = 0$; for all $c_{n}^m\in (0,1]$, $u_{n}^m\in (0,P_n^{\text{max}}]$, the following relationship can be observed:
\begin{equation}
    u_{n}^m \frac{\partial L}{\partial u_{n}^m } + c_{n}^m \frac{\partial L}{\partial c_{n}^m } \ge 0.
\end{equation}
From (\ref{s_equation}) and (\ref{c_equation}), the optimal power and subcarrier allocation scheme are obtained as:
\begin{equation} \label{optimal p}
    (P_{n}^m )^* = \frac{B\left(\lambda_n +\gamma_n E_{n}^{\text{(cm)}}\right)}{\ln2\left(\mu_n + \gamma_n \Pi\right)} - \frac{1}{\varphi_{n}^m},
\end{equation}
and 
\begin{equation} \label{optimal c}
    (c_{n}^m)^* = \begin{cases}
        1,   \qquad  {\rm if} \ \nu_n >V_{n}^m, \\
        0,  \qquad  {\rm if} \ \nu_n \leq V_{n}^m,
    \end{cases}
\end{equation}
with
\begin{equation} \label{optimal h}
    V_{n}^m =  {(R_{n}^m)}^* - \frac{1}{\ln2\left(1+\frac{1}{\varphi_{n}^m (P_{n}^m )^*}\right)},
\end{equation}
where ${(R_{n}^m)}^* = \log_2(1+\varphi_{n}^m (P_{n}^m )^*)$.
    As specified in (\ref{initial problem}e), each subcarrier can only be assigned to one user. Therefore, during the subcarrier allocation process, subcarrier $m$ is allocated to the user with the highest $V_{n}^m$. In cases where multiple users have the same maximum $V_{n}^m$, the subcarrier is assigned to the user with the highest path loss.
To find the optimal $t$, the following problem need to be considered:
\begin{equation}
    \begin{split}
        \textbf{P7}: \min \quad &\ \ t \\
        \text{s.t.} \ \ \  & \left(t_{n} ^\text{(cp)}\right)^*+\frac{\Pi}{\Tilde{R}_n} \leq t,  \quad \forall n .
    \end{split}
\end{equation}
It can be concluded that:
\begin{equation}
    t^* = \begin{cases}
        0,   \qquad \qquad \qquad \quad \quad \ \  { \text{if}} \ \sum_{n=1}^N \lambda_n \ge B\Pi, \\
        \max\limits_{n} \left\{ \left(t_{n} ^\text{(cp)}\right)^*+\frac{\Pi}{\Tilde{R}_n} \right\},   \qquad\quad\,     {\text{otherwise}}. 
    \end{cases}
\end{equation}
Furthermore, the optimal Lagrange multipliers can be calculated by solving the dual problem. Since the dual problem is a nondifferentiable convex optimization problem, the subgradient algorithm  is employed to iteratively search the optimal Lagrangian multipliers with a nonsummable diminishing stepsize $\varrho_t=\frac{0.1}{\sqrt{t}}$ \cite{boyd2003subgradient}, which are shown as:
\begin{equation} \label{update lamda}
     \lambda_n(\iota+1) =  \lambda_n(\iota) - \varrho_t \Delta \lambda_n(\iota),
\end{equation}
\begin{equation} \label{update gamma}
    \gamma_n(\iota+1) =  \gamma_n(\iota) - \varrho_t \Delta \gamma_n(\iota),
\end{equation}
\begin{equation} \label{update mu}
    \mu_n(\iota+1) =  \mu_n(\iota) - \varrho_t \Delta \mu_k(\iota),
\end{equation}
where $\iota$ is the iteration step index, and 
\begin{equation}
    \Delta \lambda_n = \frac{\Pi}{t-\left( t_n^{\text{(cp)}}\right)^*} - \Tilde{R}_{n},
\end{equation}
\begin{equation}
    \Delta \gamma_n = \sum\nolimits_{m=1}^{M}  u_{n}^m  \Pi - E_{n}^{\text{(cm)}} \Tilde{R}_{n},
\end{equation}
\begin{equation} \label{delta mu}
    \Delta \mu_n = \sum\nolimits_{m=1}^{M}u_{n}^m - P_n^{\text{max}},
\end{equation}
are the subgardients for each user, respectively. By iteratively solving the relaxed Lagrangian problem and the dual problem, LDRA can converge to a near-optimal solution that satisfies the constraints of the original optimization problem \textbf{P2-2}.
\end{IEEEproof}

\section{} \label{appendixD}
\begin{IEEEproof}Motivated by the optimal power allocation calculated through K.K.T. condition in (\ref{optimal p}), the power allocated to all the subcarriers allocated to user $n$ is proportional to their subcarrier quality and can be achieved through water-filling algorithm with a fixed water line $\theta_n^\text{WL}$. Since the transmission rate is monotonically increasing with respect to the power (or $\theta_n^\text{WL}$), the feasible $\theta_n^\text{WL}$ can be calculated through (\ref{initial problem}c) and (\ref{local communication resource allocation}a), which is given by:
\begin{equation}
    \theta_n^{\text{P}} = \frac{P_n^\text{max}+\sum\limits_{m\in\mathcal{U}_n}\frac{1}{\varphi_{n}^m}}{|\mathcal{U}_n|},
\end{equation}
\begin{equation}
    \theta_n^{\text{E}} = -\frac{W({-a_n^\text{L}} 2^{b_n^\text{L}}\ln 2)}{a_l \ln2},
\end{equation}
where $a_{n}^\text{L}\! =\! \frac{\Pi}{B E_n^\text{(cm)}}$ and $b_{n}^\text{L}\!  =\! -\frac{B E_n^\text{(cm)} \sum\limits_{m\in\mathcal{S}_n}\log_2\varphi_{n}^m +\sum\limits_{m\in\mathcal{S}_n}\frac{\Pi}{\varphi_{n}^m}}{|\mathcal{S}_n|B E_n^\text{(cm)}}$.
Then $\theta_n^\text{WL}$ can be chosen as the feasible value which is given in (\ref{lowcomplexity power allocation}). This completes the proof.
\end{IEEEproof}



\ifCLASSOPTIONcaptionsoff
  \newpage
\fi



%
\normalem
\bibliographystyle{IEEEtran}

\bibliography{IEEEabrv, IEEEexample}

\begin{thebibliography}{10}
\providecommand{\url}[1]{#1}
\csname url@samestyle\endcsname
\providecommand{\newblock}{\relax}
\providecommand{\bibinfo}[2]{#2}
\providecommand{\BIBentrySTDinterwordspacing}{\spaceskip=0pt\relax}
\providecommand{\BIBentryALTinterwordstretchfactor}{4}
\providecommand{\BIBentryALTinterwordspacing}{\spaceskip=\fontdimen2\font plus
\BIBentryALTinterwordstretchfactor\fontdimen3\font minus \fontdimen4\font\relax}
\providecommand{\BIBforeignlanguage}[2]{{%
\expandafter\ifx\csname l@#1\endcsname\relax
\typeout{** WARNING: IEEEtran.bst: No hyphenation pattern has been}%
\typeout{** loaded for the language `#1'. Using the pattern for}%
\typeout{** the default language instead.}%
\else
\language=\csname l@#1\endcsname
\fi
#2}}
\providecommand{\BIBdecl}{\relax}
\BIBdecl

\bibitem{wuchangxiang}
C.~Wu, Y.~Ren, and D.~So, ``Adaptive user scheduling and resource allocation in wireless federated learning networks : A deep reinforcement learning approach,'' in \emph{Proc. IEEE Int. Conf. Commun. (ICC)}, 2023, pp. 1231--1237.

\bibitem{jordan2015machine}
M.~I. Jordan and T.~M. Mitchell, ``Machine learning: Trends, perspectives, and prospects,'' \emph{Science}, vol. 349, no. 6245, pp. 255--260, 2015.

\bibitem{rydning2018digitization}
D.~R.-J. G.-J. Rydning, J.~Reinsel, and J.~Gantz, ``The digitization of the world from edge to core,'' \emph{Framingham: Int. Data Corp.}, 2018.

\bibitem{8016573}
Y.~Mao, C.~You, J.~Zhang, K.~Huang, and K.~B. Letaief, ``A survey on mobile edge computing: The communication perspective,'' \emph{{IEEE} Commun. Surveys Tuts.}, vol.~19, no.~4, pp. 2322--2358, 2017.

\bibitem{mcmahan2017communication}
B.~McMahan, E.~Moore, D.~Ramage, S.~Hampson, and B.~A. y~Arcas, ``Communication-efficient learning of deep networks from decentralized data,'' in \emph{Proc. 20th Int. Conf. Artificial Intelligence and Statistics}.\hskip 1em plus 0.5em minus 0.4em\relax PMLR, 2017, pp. 1273--1282.

\bibitem{li2019convergence}
X.~Li, K.~Huang, W.~Yang, S.~Wang, and Z.~Zhang, ``On the convergence of fedavg on {N}on-iid data,'' in \emph{Proc. Int. Conf. Learning Representations (ICLR)}, 2019.

\bibitem{9084352}
T.~Li, A.~K. Sahu, A.~Talwalkar, and V.~Smith, ``Federated learning: Challenges, methods, and future directions,'' \emph{{IEEE} Signal Process. Mag.}, vol.~37, no.~3, pp. 50--60, 2020.

\bibitem{pmlr-v80-katharopoulos18a}
A.~Katharopoulos and F.~Fleuret, ``Not all samples are created equal: Deep learning with importance sampling,'' in \emph{Proc. Int. Conf. Mach. Learn. (ICML)}.\hskip 1em plus 0.5em minus 0.4em\relax PMLR, 2018, pp. 2525--2534.

\bibitem{10197174}
L.~Fu, H.~Zhang, G.~Gao, M.~Zhang, and X.~Liu, ``Client selection in federated learning: Principles, challenges, and opportunities,'' \emph{{IEEE} Internet Things J.}, vol.~10, no.~24, pp. 21\,811--21\,819, 2023.

\bibitem{9904868}
E.~Rizk, S.~Vlaski, and A.~H. Sayed, ``Federated learning under importance sampling,'' \emph{{IEEE} Trans. Signal Process.}, pp. 5381--5396, 2022.

\bibitem{10443546}
B.~Luo, W.~Xiao, S.~Wang, J.~Huang, and L.~Tassiulas, ``Adaptive heterogeneous client sampling for federated learning over wireless networks,'' \emph{{IEEE} Trans. Mobile Comput.}, pp. 9663--9677, 2024.

\bibitem{10525198}
Z.~Zhu, Y.~Shi, P.~Fan, C.~Peng, and K.~B. Letaief, ``{ISFL}: Federated learning for {N}on-i.i.d. data with local importance sampling,'' \emph{{IEEE} Internet Things J.}, vol.~11, no.~16, pp. 27\,448--27\,462, 2024.

\bibitem{pmlr-v151-jee-cho22a}
Y.~Jee~Cho, J.~Wang, and G.~Joshi, ``Towards understanding biased client selection in federated learning,'' in \emph{Proc. 25th Int. Conf. Artificial Intelligence and Statistics}.\hskip 1em plus 0.5em minus 0.4em\relax PMLR, 2022, pp. 10\,351--10\,375.

\bibitem{273723}
F.~Lai, X.~Zhu, H.~V. Madhyastha, and M.~Chowdhury, ``Oort: Efficient federated learning via guided participant selection,'' in \emph{Proc. USENIX Symp. Operat. Syst. Design Implement. (OSDI)}, 2021, pp. 19--35.

\bibitem{chenmingzheconvergencetime}
M.~Chen, H.~V. Poor, W.~Saad, and S.~Cui, ``Convergence time optimization for federated learning over wireless networks,'' \emph{{IEEE} Trans. Wireless Commun.}, vol.~20, no.~4, pp. 2457--2471, 2021.

\bibitem{hard2018federated}
A.~Hard \emph{et~al.}, ``Federated learning for mobile keyboard prediction,'' \emph{arXiv preprint arXiv:1811.03604}, 2018.

\bibitem{li2020review}
L.~Li, Y.~Fan, M.~Tse, and K.-Y. Lin, ``A review of applications in federated learning,'' \emph{Comput. \& Ind. Engr.}, vol. 149, p. 106854, 2020.

\bibitem{xujie}
J.~Xu and H.~Wang, ``Client selection and bandwidth allocation in wireless federated learning networks: A long-term perspective,'' \emph{{IEEE} Trans. Wireless Commun.}, vol.~20, no.~2, pp. 1188--1200, 2021.

\bibitem{wangshiqiangfederated}
S.~Wang, T.~Tuor, T.~Salonidis, K.~K. Leung, C.~Makaya, T.~He, and K.~Chan, ``Adaptive federated learning in resource constrained edge computing systems,'' \emph{{IEEE} J. Sel. Areas Commun.}, vol.~37, no.~6, pp. 1205--1221, 2019.

\bibitem{wanshuo}
S.~Wan, J.~Lu, P.~Fan, Y.~Shao, C.~Peng, and K.~B. Letaief, ``Convergence analysis and system design for federated learning over wireless networks,'' \emph{{IEEE} J. Sel. Areas Commun.}, vol.~39, pp. 3622--3639, 2021.

\bibitem{yangzhaohui}
Z.~Yang, M.~Chen, W.~Saad, C.~S. Hong, and M.~Shikh-Bahaei, ``Energy efficient federated learning over wireless communication networks,'' \emph{{IEEE} Trans. Wireless Commun.}, vol.~20, no.~3, pp. 1935--1949, 2021.

\bibitem{yaojingjing}
J.~Yao and N.~Ansari, ``Enhancing federated learning in fog-aided iot by cpu frequency and wireless power control,'' \emph{{IEEE} Internet Things J.}, vol.~8, no.~5, pp. 3438--3445, 2021.

\bibitem{Nishio}
T.~Nishio and R.~Yonetani, ``Client selection for federated learning with heterogeneous resources in mobile edge,'' in \emph{Proc. IEEE Int. Conf. Commun. (ICC)}, 2019, pp. 1--7.

\bibitem{shiwenqi}
W.~Shi, S.~Zhou, Z.~Niu, M.~Jiang, and L.~Geng, ``Joint device scheduling and resource allocation for latency constrained wireless federated learning,'' \emph{{IEEE} Trans. Wireless Commun.}, vol.~20, pp. 453--467, 2020.

\bibitem{chenmingzhejointlearning}
M.~Chen, Z.~Yang, W.~Saad, C.~Yin, H.~V. Poor, and S.~Cui, ``A joint learning and communications framework for federated learning over wireless networks,'' \emph{{IEEE} Trans. Wireless Commun.}, vol.~20, no.~1, pp. 269--283, 2021.

\bibitem{luobing}
B.~Luo, X.~Li, S.~Wang, J.~Huang, and L.~Tassiulas, ``Cost-effective federated learning in mobile edge networks,'' \emph{{IEEE} J. Sel. Areas Commun.}, vol.~39, no.~12, pp. 3606--3621, 2021.

\bibitem{needell2014stochastic}
D.~Needell, R.~Ward, and N.~Srebro, ``Stochastic gradient descent, weighted sampling, and the randomized kaczmarz algorithm,'' \emph{Proc. Adv. Neural Inf. Process. Syst.}, pp. 1017--1025, 2014.

\bibitem{zhangjie}
J.~Zhang, S.~Guo, Z.~Qu, D.~Zeng, Y.~Zhan, Q.~Liu, and R.~Akerkar, ``Adaptive federated learning on {N}on-iid data with resource constraint,'' \emph{{IEEE} Trans. Comput.}, vol.~71, no.~7, pp. 1655--1667, 2022.

\bibitem{wanghao}
H.~Wang, Z.~Kaplan, D.~Niu, and B.~Li, ``Optimizing federated learning on non-iid data with reinforcement learning,'' in \emph{Proc. IEEE Conf. Comput. Commun. (IEEE INFOCOM)}, 2020, pp. 1698--1707.

\bibitem{zhanghangjia}
H.~Zhang, Z.~Xie, R.~Zarei, T.~Wu, and K.~Chen, ``Adaptive client selection in resource constrained federated learning systems: A deep reinforcement learning approach,'' \emph{{IEEE} Access}, vol.~9, pp. 98\,423--98\,432, 2021.

\bibitem{zhanyufengl4l}
Y.~Zhan, P.~Li, L.~Wu, and S.~Guo, ``{L4L}: Experience-driven computational resource control in federated learning,'' \emph{{IEEE} Trans. Comput.}, vol.~71, no.~4, pp. 971--983, 2022.

\bibitem{yinbenshun}
B.~Yin, Z.~Chen, and M.~Tao, ``Joint user scheduling and resource allocation for federated learning over wireless networks,'' in \emph{IEEE Glob. Commun. Conf. (GLOBECOM)}, 2020, pp. 1--6.

\bibitem{6951347}
X.~Lu, P.~Wang, D.~Niyato, D.~I. Kim, and Z.~Han, ``Wireless networks with {RF} energy harvesting: A contemporary survey,'' \emph{{IEEE} Commun. Surveys Tuts.}, vol.~17, no.~2, pp. 757--789, 2015.

\bibitem{9556559}
X.~Zhang, M.~Hong, S.~Dhople, W.~Yin, and Y.~Liu, ``Fed{PD}: A federated learning framework with adaptivity to {N}on-{IID} data,'' \emph{{IEEE} Trans. Signal Process.}, vol.~69, pp. 6055--6070, 2021.

\bibitem{9759241}
C.~Feng, H.~H. Yang, D.~Hu, Z.~Zhao, T.~Q.~S. Quek, and G.~Min, ``Mobility-aware cluster federated learning in hierarchical wireless networks,'' \emph{{IEEE} Trans. Wireless Commun.}, vol.~21, pp. 8441--8458, 2022.

\bibitem{PPO}
J.~Schulman, F.~Wolski, P.~Dhariwal, A.~Radford, and O.~Klimov, ``Proximal policy optimization algorithms,'' \emph{arXiv:1707.06347}, 2017.

\bibitem{zhao2018federated}
Y.~Zhao, M.~Li, L.~Lai, N.~Suda, D.~Civin, and V.~Chandra, ``Federated learning with non-iid data,'' \emph{arXiv preprint arXiv:1806.00582}, 2018.

\bibitem{8303773}
S.~Wang, H.~Liu, P.~H. Gomes, and B.~Krishnamachari, ``Deep reinforcement learning for dynamic multichannel access in wireless networks,'' \emph{{IEEE} Trans. on Cogn. Commun. Netw.}, vol.~4, no.~2, pp. 257--265, 2018.

\bibitem{bertsekas1997nonlinear}
D.~P. Bertsekas, ``Nonlinear programming,'' \emph{Journal of the Operational Research Society}, no.~3, pp. 334--334, 1997.

\bibitem{793310}
C.~Y. Wong, R.~Cheng, K.~Lataief, and R.~Murch, ``Multiuser ofdm with adaptive subcarrier, bit, and power allocation,'' \emph{{IEEE} J. Sel. Areas Commun.}, vol.~17, no.~10, pp. 1747--1758, 1999.

\bibitem{chou2017improving}
P.-W. Chou, D.~Maturana, and S.~Scherer, ``Improving stochastic policy gradients in continuous control with deep reinforcement learning using the beta distribution,'' in \emph{Int. Conf. on Mach. Learn. (ICML)}.\hskip 1em plus 0.5em minus 0.4em\relax PMLR, 2017, pp. 834--843.

\bibitem{boyd2003subgradient}
S.~Boyd, L.~Xiao, and A.~Mutapcic, ``Subgradient methods,'' \emph{lecture notes of EE392o, Stanford University, Autumn Quarter}, 2003.

\end{thebibliography}




%








\end{document}